\documentclass[twocolumn,trackchanges]{aastex631}

\usepackage{graphicx} 
\usepackage{amsmath}
\usepackage{subfigure}
\usepackage{makecell}
\usepackage{url}
\usepackage{longtable}

\begin{document}
\title{Key Physical Parameters Influencing Fragmentation and Multiplicity in Dense Cores of Orion~A}

\author[0009-0002-4564-7679]{Jo-Shui Kao}
\affiliation{Institute of Astronomy, National Tsing Hua University, No. 101, Section 2, Kuang-Fu Rd., Hsinchu 30013, Taiwan}
\affiliation{Academia Sinica Institute of Astronomy and Astrophysics,
11F of Astronomy-Mathematics Building, AS/NTU, No.1, Sec. 4, Roosevelt Rd, Taipei 10617, Taiwan}

\author[0000-0003-1412-893X]{Hsi-Wei Yen}
\affiliation{Academia Sinica Institute of Astronomy and Astrophysics,
11F of Astronomy-Mathematics Building, AS/NTU, No.1, Sec. 4, Roosevelt Rd, Taipei 10617, Taiwan}

\author[0000-0001-5522-486X]{Shih-Ping Lai}

\affiliation{Institute of Astronomy, National Tsing Hua University, No. 101, Section 2, Kuang-Fu Rd., Hsinchu 30013, Taiwan}
\affiliation{Academia Sinica Institute of Astronomy and Astrophysics,
11F of Astronomy-Mathematics Building, AS/NTU, No.1, Sec. 4, Roosevelt Rd, Taipei 10617, Taiwan}
\affiliation{Center for Informatics and Computation in Astronomy, National Tsing Hua University, No. 101, Section 2, Kuang-Fu Rd., Hsinchu 30013, Taiwan}
\affiliation{Department of Physics, National Tsing Hua University, No. 101, Section 2, Kuang-Fu Rd., Hsinchu 30013, Taiwan}



\begin{abstract}
When dense cores in molecular clouds or filamentary structures collapse and form protostars, they may undergo fragmentation and form binary or multiple systems.
In this paper, we investigated the key mechanisms influencing fragmentation by comparing the physical conditions of fragmented and unfragmented dense cores ($\sim$0.1 pc) in Orion A. 
Utilizing archival submillimeter continuum data from the James Clerk Maxwell Telescope (JCMT) and the Atacama Large Millimeter/Submillimeter Array survey of Class 0 and I protostars at a 0\farcs1 resolution, we identified 38 dense cores hosting single protostars and 15 cores hosting binary or multiple systems. 
We measured the dense cores properties with the \textit{Herschel} dust temperature, Nobeyama 45m N$_2$H$^+$ J=1-0, and JCMT polarization data. 
Our results reveal that the dense cores hosting binary/multiple systems exhibit significantly higher density and Mach number compared to those hosting single protostars, while there are no correlations between the occurrence of fragmentation and the energy ratios of turbulence and magnetic field to gravity. 
Our results suggest that the higher density and supersonic turbulence of the dense cores can lead to local collapse and fragmentation to form binary/multiple systems, while the magnetic field has limited influence on fragmentation in the dense cores in Orion A.
\end{abstract}



\section{Introduction} \label{sec:intro}
\par
Main-sequence stars are frequently found in binary or multiple systems \citep{2013ARA&A..51..269D}.
Previous studies on the multiplicity and companion fraction of protostars show that a significant fraction of stars are born within binary or multiple systems during the early stages of star formation, and eventually evolve to be main-sequence stars \citep{2007A&A...476..229D,2008AJ....135.2526C,2016ApJ...821...52K,2022ApJ...925...39T, 2023ASPC..534..275O}.

\par
Turbulence within dense cores is theoretically expected to play a crucial role in the formation process of binary or multiple systems \citep{2004MNRAS.351..617D, 2018A&A...615A..94F}.
The turbulent core fragmentation scenario suggests that turbulent fluctuations within a collapsing core can lead to the emergence of multiple density peaks.
If the local mass of these peaks surpasses the local Jeans mass, they will collapse faster than the bound core and form multiple protostars \citep{2004A&A...414..633G, 2007prpl.conf..133G,2010ApJ...725.1485O}.
Theoretical simulations show that the companion frequency increases as the ratio of turbulent to gravitational energy increases \citep{2004A&A...423..169G}.
On the other hand, besides prompting local collapse, turbulence can provide extra support against gravitational collapse.
This could lower the star formation efficiency and influence the multiplicity \citep{2004MNRAS.347L..36C, 2007ARA&A..45..565M, 2007prpl.conf...63B, 2012A&ARv..20...55H, 2024MNRAS.528.1460R}.

\par
In addition, dense cores are magnetized \citep{2019FrASS...6....5H, 2023ASPC..534..193P}.
Theoretically, the magnetic field can delay or suppress fragmentation \citep{2011ApJ...729...72P, 2020SSRv..216...70L, 2022FrASS...9.9223M}.
Numerical studies show that the core rotation can enhance fragmentation, while the magnetic braking can remove the angular momentum from the central part \citep{2004MNRAS.347.1001H, 2008ApJ...677..327M, 2010A&A...510L...3C}.
The extra support from the magnetic pressure can reduce the infall rate of mass from the envelope to the disk, keeping it from gravitational instabilities \citep{2007MNRAS.377...77P, 2013ApJ...766...97M, 2018MNRAS.480.3511G}.
Both effects can suppress fragmentation.
On the contrary, the magnetic field may enhance formation of multiple systems by inducing bar or filamentary structure since dense cores more easily collapse along the magnetic field \citep{1982A&A...114..151D,1984A&A...139..378B,2019ApJ...878...10T}.
These elongated structures could be more unstable and prone to fragment \citep{2000ApJ...545L..61B,2000ApJ...528..325B}.

\par
Observationally, the influence of turbulence and magnetic field on the fragmentation process is still under debate.
The observations of hierarchical fragmentation from clump (1 pc), core (0.1 pc) to condensation (0.01 pc) scales in the infrared dark cloud found that the masses of the cores and condensations are comparable to the turbulent Jeans mass, suggesting that the contribution of turbulent support is important in the fragmentation process \citep{2014MNRAS.439.3275W}.
In contrast, the other study toward the nearby massive dense cores of ten to a few tens of solar mass shows a clear correlation between the number of fragments and the ratio of the core mass to thermal Jeans mass.
This suggests that the thermal Jeans fragmentation is the dominant process of binary or multiple formation within these dense cores \citep{2015MNRAS.453.3785P}.

The observations of magnetic fields and core fragmentation on a 0.3 pc scale in the two hub–filament systems have found a stronger magnetic field in the less fragmented hub, supporting that the magnetic field can suppress core fragmentation \citep{2016ApJ...819..139B, 2020A&A...644A..52A}.
However, such a relation is not seen in the other dense clumps ($\sim$ 1 pc) in 20 high-mass star-forming regions, where there is no clear correlation between the number of fragments and the magnetic field strengths \citep{2024A&A...682A..81B}.

These discrepancies could be due to the relative importance between turbulence, magnetic fields, and gravity, which could play an important role in the fragmentation process and may affect the fragmentation type \citep{2019ApJ...878...10T}.
Therefore, to investigate the key physcial parameters influencing core fragmentation and multiplicity, observations assessing the energy ratio of the turbulence and magnetic fields to gravity in a large sample of fragmented and unfragmented dense cores spanning a wide range of physical conditions are required.



\par
Orion A is one of the nearest star-forming regions, where the multiplicity has been measured with the VLA/ALMA Nascent Disk And Multiplicity (VANDAM) survey \citep{2020ApJ...890..130T}.
Its molecular-line and polarized thermal dust emission at (sub-)millimeter wavelengths have been mapped over a wide area of more than 1$\fdg$5 by 0$\fdg$5 \citep{2007PASP..119..855W,2008PASJ...60..407T,2017ApJ...846..122P}. This enables the measurement of the turbulent and magnetic energy in a large sample of dense cores in Orion~A. Therefore, it is an ideal region to study core fragmentation in comparison with their physical conditions.
\cite{2022ApJ...931..158L} have studied 43 protostellar cores inside Planck Galactic Cold Clumps (PGCCs) in the Orion molecular cloud complex and found that the cores with higher gas density and Mach number tend to form binary or multiple system. To further extend this study, in our work, we analyze the turbulence and magnetic fields in a complementary sample of 38 unfragmented and 15 fragmented dense cores in Orion A.

\par
In this paper, we present the physical conditions of the dense cores forming binary or multiple systems and compare to those forming single protostars to investigate the key mechanisms of fragmentation in dense cores in Orion A.
In Section \ref{sec:data}, we introduce the observational data adopted in this paper.
Section \ref{sec:analysis} describes our analysis of dense core properties.
Section \ref{sec:results} presents the comparisons of the properties between single and binary/multiple systems.
Section \ref{sec:discussion} discusses the possible mechanisms of dense core fragmentation in Orion A in comparison with other star-forming regions.
Finally, we summarize this work in section \ref{sec:summary}.

\section{Data} \label{sec:data}
To assess the turbulence, magnetic fields, and gravity in the dense cores in Orion A, we made use of the James Clerk Maxwell Telescope (JCMT) SCUBA-2/POL-2 data, the \textit{Herschel} dust temperature map, and the Nobeyama 45-m N$_2$H$^+$ J = 1--0 data, and we adopted the catalog of protostars in Orion A from the VANDAM Survey.

\subsection{Protostar Sample}
To assess the multiplicity of dense cores, we retrieved the position information of protostars from the catalog of the VANDAM Survey \citep[Table 6 in][]{2020ApJ...890..130T}. The VANDAM survey observed 328 Class 0, Class I, and flat-spectrum protostars in the Orion molecular clouds with the Atacama Large Millimeter/submillimeter Array (ALMA) at 0.87 mm and a resolution of $\sim 0\farcs1$ (40 au), and it is sensitive to the projected separations larger than 20 au. 
In this sample, there are 275 protostars in the Orion~A region.

\subsection{JCMT Data}
We retrieved the JCMT POL-2 data of the Orion A molecular clouds taken under the projects M16BD001, M17BL011, M20AL006, M20AL018, M21BF005, and M21BF004 from the public archive.
These observations were conducted during 2017 to 2023. 
Part of these data has been presented in \cite{2017ApJ...846..122P}.
In addition, we conducted new observations toward Orion A from 2024 January 22 to 28 with our PI project M23BP064. 
The observations were carried out with the Band 1 and 2 weather conditions and the POL-2 daisy observing mode. 
The point center was 05$^{\rm h}$36$^{\rm m}$00, $-$06$^\circ$23$^\prime$00, and the total observing time was 7.7 hours.

For the present paper, we used the POL-2 data at 850 $\mu$m. 
The beam size of the JCMT observations at 850 $\mu$$m$ is 14$\farcs$6.
All the POL-2 850 $\mu$m data were calibrated together using the software {\it starlink} version 2023A \citep{2014ASPC..485..391C, 2022ASPC..532..559B} and its task {\it pol2map}. 
The default pixel size of 4$\arcsec$ was adopted in the calibration process. 
The calibrated Stokes $IQU$ maps were binned to a pixel size of 12$\arcsec$, approximately the JCMT beam size, to extract polarization detection, and the polarized intensity was debiased.

The polarization detections were extracted from the binned Stokes $IQU$ maps when the signal-to-noise ratios of Stokes $I$ is above 5, signal-to-noise ratios of polarization intensities is above 2, and polarization percentage is below 30\% in the pixels. 
Then the magnetic field orientations were inferred by rotating the polarization orientations by $90^\circ$. 
The uncertainties of the polarization orientations and thus the uncertainties of magnetic field orientations range from less than $1^\circ$ to $16^\circ$. 
The noise level of the Stokes $I$ map is estimated to be 10 mJy.
The resultant JCMT maps are presented in Figure \ref{fig:map}, \ref{fig:map_in}, and Appendix \ref{app:map} (Figure \ref{fig:core}).

\subsection{Herschel Dust Temperature Map}
We retrieved the dust temperature map of Orion A from the public data archive of the \textit{Herschel} Gould Belt Survey \citep[HGBS;][]{2010A&A...518L.102A,2013ApJ...763...55R,2013ApJ...777L..33P}, which have been presented in \citet{2013ApJ...763...55R} and \citet{2013ApJ...777L..33P}.
The dust temperature was estimated by fitting the spectral energy distributions at 160, 250, 350, and 500 $\mu$m. 
The \textit{Herschel} dust temperature map has a pixel size of 4$\arcsec$ and an angular resolution of $37\arcsec$.
The retrieved Herschel dust temperature map is presented in Appendix \ref{app:map} (Figure \ref{fig:temp_map}).

\subsection{Nobeyama 45-m N$_2$H$^+$ J=1--0}\label{sec:N2H+}
We obtained the N$_2$H$^+$ J=1--0 image cube taken by the 45-m telescope of the Nobeyama Radio Observatory from \citet{2008PASJ...60..407T}. 
The data and observations have been presented in detail in \citet{2008PASJ...60..407T}. The N$_2$H$^+$ (1--0) data has a pixel size of $7.5\arcsec$, a beam size of 23$\farcs$4, and a spectral resolution of 37.8 kHz, corresponding to 0.12 km s$^{-1}$. 
The N$_2$H$^+$ (1--0) observations mapped the whole $\int$-shape filament in Orion A with declination ranging from $-$04$^\circ$48$^\prime$05 to $-$07$^\circ$11$^\prime$51, covering the entire JCMT POL-2 field. 
The examples of the obtained N$_2$H$^+$ (1--0) maps are presented in Appendix \ref{app:map}.

\section{Analysis} \label{sec:analysis}
To investigate the core fragmentation in Orion A, we identified the dense cores in the JCMT 850 $\mu$m map, and analyze their physical conditions, including mass density, Jeans length, Mach number, turbulent energy, magnetic field strength, and mass-to-flux ratio. 
In this section, we introduce our analysis, and our derived core properties are presented in Table \ref{tab:core} and \ref{tab:B}. 
Their uncertainties are estimated from the error propagation. 
We note that that the resolutions of these data sets from JCMT, \textit{Herschel}, and Nobeyama 45-m telescope are not identical. 
Nevertheless, we focus on the mean physical parameters averaging over the size of the dense cores, which is typically larger than or comparable to the resolutions, and thus its impact is minimal. 

\subsection{Core Identification} \label{sec:core}

\begin{figure*}[htbp]
\centering
\includegraphics[width=0.9\textwidth]{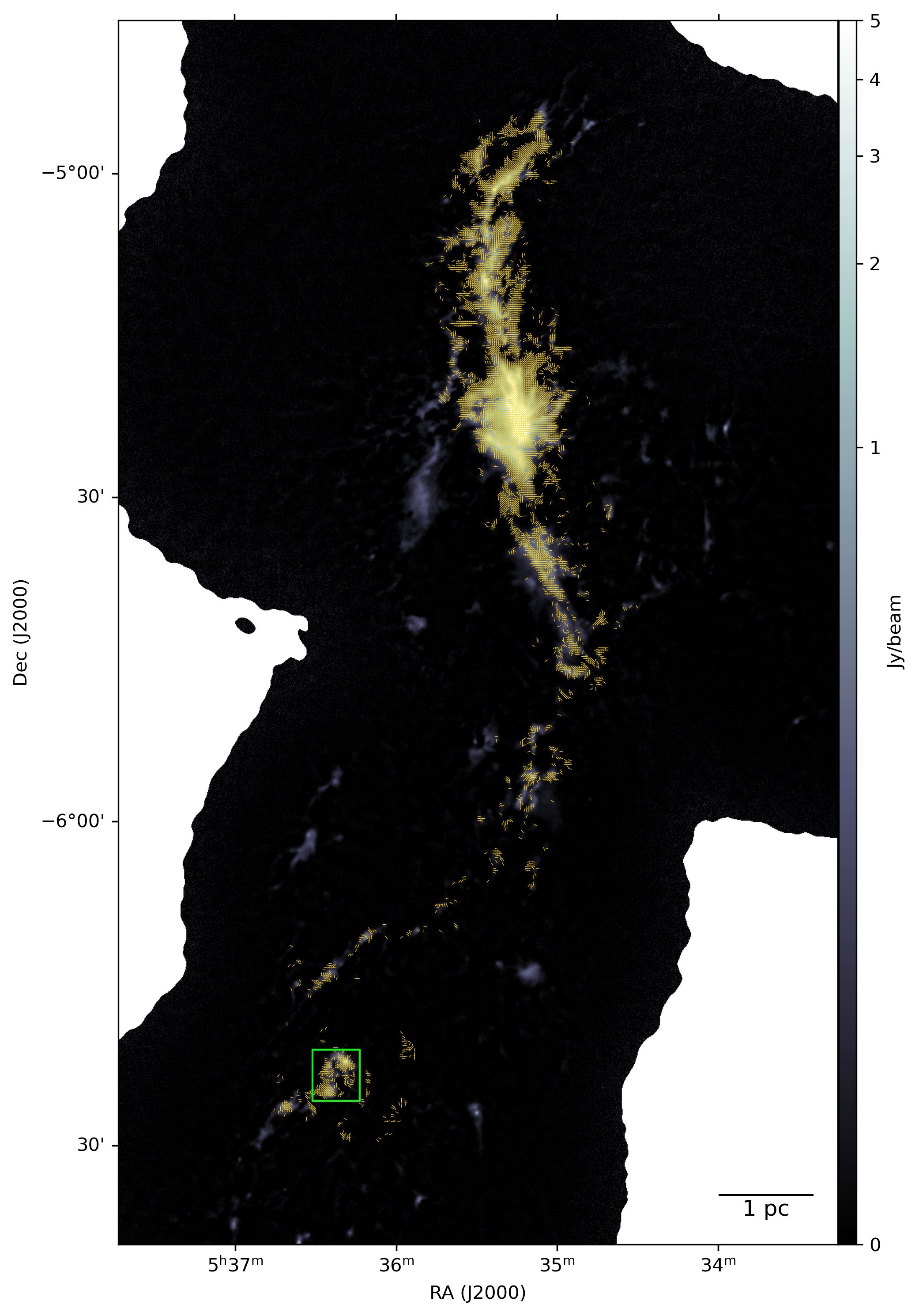}
\caption{850 $\mu$m continuum map observed with JCMT. 
The blank area is outside the field of views of the observations. 
The yellow segments show the magnetic field orientations obtained by rotating the observed polarization orientations by $90^\circ$. 
The pixel size of the polarization map is $12\,\arcsec$, and there are 5523 polarization detections in total in this map. 
The zoomed-in view of the green box is shown in the following figure.}
\label{fig:map}
\end{figure*}

\begin{figure}[htbp]
    \centering
    \includegraphics[width=0.48\textwidth]{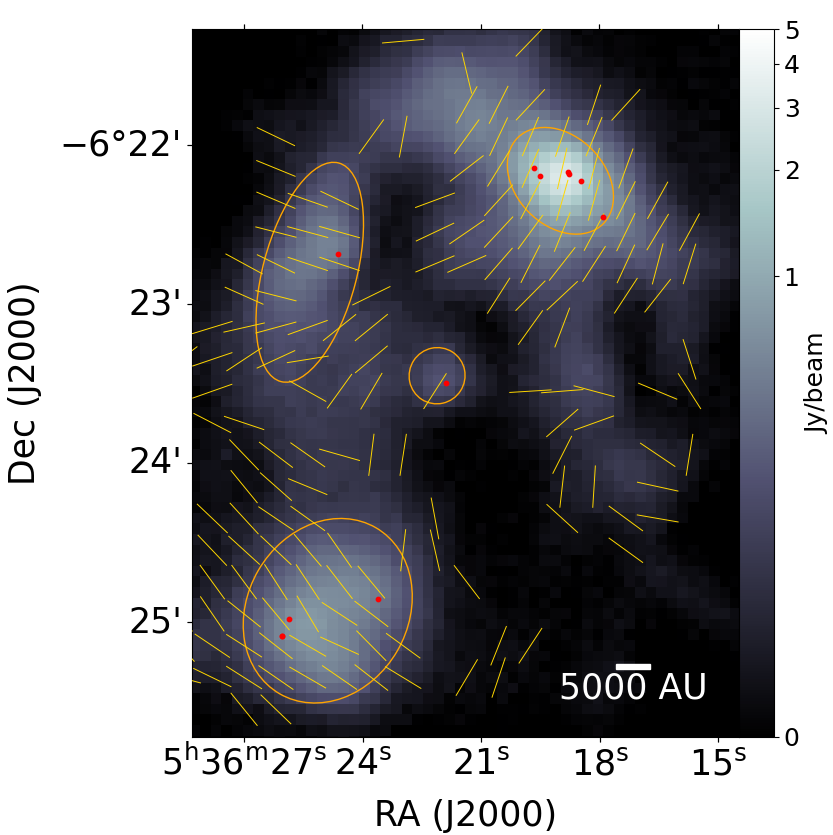}
    \caption{Zoomed-in view to the green box in Figure \ref{fig:map}, which is obtained with our PI JCMT observations. 
    The yellow segments show the magnetic field orientations obtained by rotating the observed polarization orientations by $90^\circ$. 
    The separation between the segments is 12$\arcsec$.
    The red dots mark the positions of protostars from the VANDAM survey.
    The orange ellipses label the dense cores identified by our \textit{astrodendro} analysis.
    The semimajor and semiminor axes of the ellipses are the 2$\sigma$ widths of the fitted 2D Gaussian distribution.}
    \label{fig:map_in}
\end{figure}

Figure \ref{fig:map} presents the JCMT 850 $\mu$m map.
We identified dense cores in the JCMT 850 $\mu$m map by applying the \textit{astrodendro} package which builds dendrograms of observed data and fits a 2D Gaussian function to each leaf.
In our analysis, we adopted the minimum threshold of 50 mJy/beam (5$\sigma$ noise level), minimum significance of 30 mJy/beam (3$\sigma$ noise level), and minimum number of pixels of 5, approximately half of the beam size. 
We identified 493 dense cores, meaning the leaf structures from the dendrograms, within the JCMT 850 $\mu$m map.
Their total fluxes, positions, and position angles were measured by 2D Gaussian fitting to their 850 $\mu$m emission.
The size of the dense cores is defined to be the area within the 2$\sigma$ widths of the fitted 2D Gaussian distribution, which is expected to enclose 90\% of the total flux.
The radius of the dense core is defined to be the geometric mean of 2$\sigma$ widths of the semimajor and semiminor axes with the uncertainty estimated using the bootstrap method \citep{2006PASP..118..590R}.
We note that dense core with sizes smaller than the JCMT beam size may not be fully resolved. Their sizes can be overestimated, and their properties cannot be robustly estimated. 
Thus, these dense cores are excluded from our sample.
The radii of the dense cores in our sample range from 3200 au to 16000 au.

\par
We defined a dense core to be associated with protostars if there are one or more protostars located within the elliptical area of the dense core defined by our \textit{astrodendro} analysis. The coordinates of protostars were measured with ALMA at a 0\farcs1 resolution and obtained from Table 6 in \citet{2020ApJ...890..130T}. 
In our maps, there are 53 dense cores associated with in total 91 protostars. 
Among them, there are 38 dense cores hosting single protostars, and 5, 3, 4, and 3 dense cores hosting 2, 3, 4, and 6 protostars, respectively, with separations between the companions ranging from 45 au to 15000 au.
An example of association of the dense cores and protostars is presented in Figure \ref{fig:map_in}.

\par
We note that binary or multiple systems with separations smaller than a few hundred AU may form via disk fragmentation, or through core fragmentation and then migrate to a smaller separation \citep{2016ApJ...818...73T}.
The former is not the focus of this paper.
In our sample, there are 4 systems with separations smaller than 500 au.
We have tested it and confirmed that including or excluding these sources having multiples with small separations does not affect the results of our subsequent analysis and conclusions.
On the other hand, unresolved multiple systems (if any), meaning that their separations are smaller than 20 au, are more likely formed via disk fragmentation, so they can be considered as single systems in our discussion because our focus is core fragmentation.

\subsection{Core Mass and Density} \label{sec:density}
We estimated the core mass from the 850 $\mu$m continuum emission with the following equation \citep{1983QJRAS..24..267H,2016ApJ...817..167K},
\begin{equation}
    \begin{aligned}
        M_{\rm core}    &= \frac{S_{\nu} D^2}{B_{\nu} (T_d) \kappa_{\nu}}\\
                    &= 1.06 \times S_{\nu} \times (e^{\frac{17K}{T_d}}-1) \times (D/415 \text{pc})^2\ M_\sun,
    \end{aligned}
\end{equation}\label{eq:mass}
where $S_{\nu}$ is the total flux of the dense core at 850 $\mu$m, $D$ is the distance to the core ranging from 383 to 392 pc \citep[][]{2020ApJ...890..130T}, $T_d$ is the dust temperature, $B_{\nu}$ is the Planck function at temperature $T_d$, and $\kappa_{\nu}$ is the dust opacity at 850 $\mu$m.
Here we adopted $\kappa_{\nu}$ of 0.0125 cm$^2$ g$^{-1}$ \citep{2017ApJ...836..132J},  and its uncertainty is adopted to be $32\%$  \citep{2017ApJ...841...97S}.
The dust temperature of the dense cores is obtained from the \textit{Herschel} dust temperature map by averaging the values of the pixels within the area of the dense cores. 
The uncertainty of the dust temperature of the dense cores is estimated as the standard deviation of these pixel values. 
We have compared the brightness temperature of the observed 850 $\mu$m continuum emission with the dust temperature from the Hershel maps, and found that all the dense cores in our sample have $\tau < 0.02$ at 850 $\mu$m.
Thus, our assumption of optically thin dust emission is valid.
The estimated mass of these dense cores ranges from 0.27 $M_\sun$ to 37 $M_\sun$ with a typical uncertainty of 30\% due to the noise and the uncertainties in $\kappa$ and dust temperature.

\par
We note that a part of the total flux ($S_{\nu}$) may be contributed by circumstellar disks around protostars embedded in the dense cores. 
We have compared the total fluxes of the dense cores observed with JCMT and those observed with ALMA at 0\farcs1 in \cite{2020ApJ...890..130T}. 
In 80\% of our sample, the possible disk contribution is less than 15\% of the total flux of the dense cores. 
Only two (out of 53) dense cores hosting single protostars may have disk contribution of more than 35\% of their total flux.
Therefore, the contribution from the circumstellar disks is negligible compared to the typical uncertainty in the core mass in most of the cases, and it does not affect our statistical analysis in the present paper.

\par
In addition, we note that the mass calculated here does not include the protostellar mass, and thus the total mass within the core area may be underestimated.
Its possible impact on our analyze is discussed in Section \ref{sec:density and thermal}, \ref{sec:tub}, and \ref{sec:B_ab}.
With the core radius from Section \ref{sec:core}, we then derived the core density on the assumption of the spherical geometry.
The density of the dense cores in our sample ranges from $1.5\times10^{-19}$ g cm$^{-3}$ to $10^{-17}$ g cm$^{-3}$.

\subsection{Jeans Instability}\label{sec:jeans}
Considering only thermal pressure and gravity, the Jeans length ($\lambda_{Jeans}$) and Jeans mass are defined as
\begin{equation}\label{eq:Jeans_length}
    \lambda_{Jeans} = \frac{c_s}{\sqrt{G \rho}},
\end{equation}
\begin{equation}\label{eq:Jeans_mass}
    M_{Jeans} = \frac{4 \pi}{3} \rho R^3_{Jeans},
\end{equation}
where $c_s$ is the sound speed, $\rho$ is the density in the dense cores, and $R_{Jeans}$is Jeans radii defined as $\lambda_{Jeans}/2$.
$c_s$ is estimated as $\sqrt{k_BT/m_{H_2}}$, where $T$ is the gas temperature.
Here we assumed that the gas temperature is the same as the dust temperature.
Then we calculated the Jeans length with the temperature from the \textit{Herschel} dust temperature map and the density estimated in Section \ref{sec:density}.

\subsection{Velocity Dispersion}\label{sec:N2H+ dv}
We fitted the spectrum of the seven hyperfine components of the N$_2$H$^+$ (1--0) emission to measure the centroid velocity, line width, and optical depth in each pixel with the line detection above 3$\sigma$ (Appendix \ref{app:map}). 
We assumed that all seven hyperfine components have the same excitation temperature and line width, and the line profile was assumed to be a Gaussian function. 
The rest frequencies of the hyperfine components were adopted from the Leiden Atomic Molecular Database\footnote{\url{https://home.strw.leidenuniv.nl/~moldata/}}, and their relative intensity ratios were from \citet{1992ApJ...387..417W}. 
In some regions, two velocity components are present. 
In those pixels, we fitted the spectra with two velocity components if the Bayesian information criterion suggests that the two velocity components provide a better fit (Appendix \ref{app:map}), and we assumed that their excitation temperatures are the same \citep{2020A&A...644A..29R}. 
Then we selected the parameters of the component with a higher optical depth for the subsequent analysis.

\subsection{Mach Number and Turbulence Pressure} \label{sec:jeans}
To evaluate the turbulence within the dense cores, we estimated their Mach number ($\mathcal{M}$), a dimensionless parameter that represents the ratio of turbulent velocity to sound speed as
\begin{equation}
    \mathcal{M} = \frac{\sigma_{NT}}{c_s},
\end{equation}
where $\sigma_{NT}$ is the non-thermal velocity dispersion, derived from
\begin{equation}
    \sigma_{NT} = \sqrt{\Delta v^2 - \frac{k_BT}{m_{\rm N_2 H^+}}},
\end{equation}
where $\Delta v$ is the 1$\sigma$ line width and $m_{\rm N_2 H^+}$ is the molecular weight of N$_2$H$^+$ (1--0).
The velocity dispersion $\Delta v$ is measured from the N$_2$H$^+$ (1--0) map with spectral fitting to its seven hyperfine components (Sec.~\ref{sec:N2H+ dv}).
We calculated the average $\Delta v$ of all the pixels within the core area weighted by their optical depths, which is proportional to column density.
The estimated Mach numbers of the dense cores range from 0.4 to 2.4.
There are 7 dense cores hosting single protostar without N$_2$H$^+$ (1--0) data so their turbulence cannot be assessed.

\par
Besides Mach number, we calculated the relative importance between turbulence and gravity by comparing turbulent pressure and gravitational energy density.
The turbulent pressure ($P_T$) can be calculated as
\begin{equation}
    P_T = \frac{3}{2}\rho(\sigma_{NT})^2,
\end{equation}
where $\rho$ is the density derived in Section \ref{sec:density}.
Assuming that the dense cores are spherical, the gravitational energy density ($u_G$) can be calculated as
\begin{equation}
    u_G = \frac{9}{20\pi} G \frac{M^2}{R^4},
\end{equation}
where $G$ is the gravitational constant, $M$ is the core mass, and $R$ is the core radius \citep{2019ApJ...878...10T}.
The ratios of $P_T/u_G$ in the dense cores in our sample range from 0.06 to 2.05.

\subsection{Magnetic Field and Mass-to-flux Ratio}\label{sec:B_lambda}
The polarized continuum emission is detected in 27 cores with 3 or more detections, including 17 single systems and 10 binary or multiple systems.
We measured the mean polarization orientation of each of these cores from the mean Stokes {\it Q} and {\it U} of all polarization detections in the core area, and then rotated the polarization orientation by 90$\arcdeg$ to infer the mean magnetic field orientation in the core. 

We estimated the magnetic field strengths in the dense cores using the David-Chandrasekhar-Fermi (DCF) method \citep{1951PhRv...81..890D, 1953ApJ...118..113C} as
\begin{equation} \label{eq:dcf}
    B_{pos} = Q \sqrt{4 \pi \rho} \frac{\sigma_{NT}}{\Delta \phi_B},
\end{equation}
where $Q$ is the numerical correction factor and is adopted to be 0.5 \citep{2001ApJ...546..980O}, and $\Delta \phi_B$ is the angular dispersion of the magnetic field orientations.
To determine the angular dispersion, we first adopted the unsharp-masking method to remove the large-scale magnetic field structures \citep{2023ASPC..534..193P}.
We smoothed the magnetic field orientation map with a kernel of 3 by 3 pixels (or 36$\arcsec$ by 36$\arcsec$), corresponding to the median core size in our sample.
Subsequently, we subtracted the smoothed map from the original map, resulting in a residual map.
Then we calculated the standard deviation of the residual angles in the core as the angular dispersion $\Delta \phi_B$.
We note that $B_{pos}$ in Equation \ref{eq:dcf} is the projected magnetic field strength on the plane of the sky, and we did not correct it for projection due to unknown inclination angles.
The magnetic field strengths of the cores range from 0.04 to 3.1 mG.
\par
The relative importance of the magnetic field and gravity can be evaluated by the mass-to-flux ratio ($\lambda$) as
\begin{equation}
    \lambda = \frac{(M/\phi)_{\rm obs}}{(M/\phi)_{\rm critical}} = 7.6 \times 10^{-21} \frac{N_{H_2}}{B},
\end{equation}
where $N_{\rm H_2}$ is the H$_2$ column density, and the $B$ is the magnetic field strength measured from the DCF method in $\mu$G. 
The H$_2$ column density is derived from the core mass from Section \ref{sec:density} and the core area projected on the plane of sky from Section \ref{sec:core}.
The mass-to-flux ratios of the cores range from 0.2 to 4.0.

\begin{longrotatetable}
    \begin{deluxetable*}{ccccccccccccc}%
    \tablecaption{Core properties}
    \tablehead{
        \colhead{index} & 
        \colhead{R.A.} & 
        \colhead{Dec.} & 
        \colhead{$N_{\mathrm{yso}}$} & 
        \colhead{Radius} & 
        \colhead{$T_d$} & 
        \colhead{Mass} & 
        \colhead{N(H$_2$)} & 
        \colhead{$\rho$} & 
        \colhead{$u_G$} &
        \colhead{$\sigma_{\mathrm{NT}}$} & 
        \colhead{$P_T$} &
        \colhead{$\mathcal{M}$} \\
        \colhead{} &
        \colhead{(J2000)} & 
        \colhead{(J2000)} & 
        \colhead{} & 
        \colhead{(au)} & 
        \colhead{(K)} & 
        \colhead{($M_{\odot}$)} & 
        \colhead{$10^{21}$ g cm$^{-2}$} & 
        \colhead{$10^{-19}$ g cm$^{-3}$} & 
        \colhead{$10^{-9}$ dyn cm$^{-2}$} &
        \colhead{(m s$^{-1}$)} & 
        \colhead{$10^{-9}$ dyn cm$^{-2}$} &
        \colhead{} \\
        \colhead{(1)} & 
        \colhead{(2)} & 
        \colhead{(3)} & 
        \colhead{(4)} & 
        \colhead{(5)} & 
        \colhead{(6)} & 
        \colhead{(7)} & 
        \colhead{(8)} & 
        \colhead{(9)} &
        \colhead{(10)} & 
        \colhead{(11)} &
        \colhead{(12)} &
        \colhead{(13)}
    }
\startdata
1 & 05:38:03.59 & -06:58:17.27 & 1 & $ 5300 \pm 600 $ & $ 13.7 \pm 0.1 $ & $ 0.5 \pm 0.2 $ & $ 14 \pm 7 $ & $ 4.3 \pm 2.4 $ & $ 0.2 \pm 0.2 $ & ... & ... & ...\\ 
2 & 05:38:40.41 & -06:58:11.38 & 1 & $ 3700 \pm 600 $ & $ 14.8 \pm 0.1 $ & $ 0.4 \pm 0.2 $ & $ 26 \pm 13 $ & $ 12.3 \pm 7.4 $ & $ 0.8 \pm 0.8 $ & ... & ... & ...\\ 
3 & 05:37:56.51 & -06:56:26.24 & 1 & $ 9000 \pm 800 $ & $ 13.8 \pm 0.2 $ & $ 2.1 \pm 0.7 $ & $ 22 \pm 9 $ & $ 4.2 \pm 1.8 $ & $ 0.5 \pm 0.4 $ & ... & ... & ...\\ 
4 & 05:37:17.64 & -06:49:43.84 & 1 & $ 5100 \pm 600 $ & $ 14.4 \pm 0.1 $ & $ 0.9 \pm 0.3 $ & $ 29 \pm 12 $ & $ 9.5 \pm 4.7 $ & $ 0.9 \pm 0.7 $ & $ 212 \pm 11$ & $ 0.6 \pm 0.3 $ & $ 0.87 \pm 0.05 $\\ 
5 & 05:37:50.65 & -06:47:10.06 & 1 & $ 6300 \pm 600 $ & $ 14.7 \pm 0.1 $ & $ 0.9 \pm 0.3 $ & $ 19 \pm 8 $ & $ 5.1 \pm 2.3 $ & $ 0.4 \pm 0.3 $ & ... & ... & ...\\ 
6 & 05:36:23.17 & -06:46:11.85 & 4 & $ 10500 \pm 700 $ & $ 14.9 \pm 0.6 $ & $ 11.4 \pm 3.7 $ & $ 87 \pm 31 $ & $ 13.9 \pm 5.4 $ & $ 8.0 \pm 5.7 $ & $ 343 \pm 4$ & $ 2.5 \pm 1.0 $ & $ 1.38 \pm 0.03 $\\ 
7 & 05:36:18.53 & -06:45:41.88 & 2 & $ 10200 \pm 600 $ & $ 15.4 \pm 1.6 $ & $ 9.5 \pm 3.5 $ & $ 76 \pm 29 $ & $ 12.7 \pm 5.1 $ & $ 6.3 \pm 4.8 $ & $ 271 \pm 2$ & $ 1.4 \pm 0.6 $ & $ 1.08 \pm 0.06 $\\ 
8 & 05:36:25.26 & -06:44:56.86 & 1 & $ 8600 \pm 700 $ & $ 14.2 \pm 0.2 $ & $ 3.5 \pm 1.1 $ & $ 40 \pm 14 $ & $ 7.6 \pm 3.1 $ & $ 1.6 \pm 1.2 $ & $ 412 \pm 4$ & $ 2.0 \pm 0.8 $ & $ 1.70 \pm 0.02 $\\ 
9 & 05:36:36.30 & -06:38:54.57 & 1 & $ 7700 \pm 700 $ & $ 15.6 \pm 0.3 $ & $ 3.6 \pm 1.2 $ & $ 51 \pm 19 $ & $ 11.4 \pm 4.8 $ & $ 2.9 \pm 2.1 $ & $ 244 \pm 12$ & $ 1.0 \pm 0.4 $ & $ 0.96 \pm 0.05 $\\ 
10 & 05:36:17.21 & -06:38:03.12 & 1 & $ 6400 \pm 600 $ & $ 15.8 \pm 0.1 $ & $ 0.9 \pm 0.3 $ & $ 19 \pm 7 $ & $ 4.9 \pm 2.2 $ & $ 0.4 \pm 0.3 $ & $ 105 \pm 3$ & $ 0.08 \pm 0.04 $ & $ 0.41 \pm 0.01 $\\ 
11 & 05:37:00.61 & -06:37:14.40 & 1 & $ 9500 \pm 600 $ & $ 13.0 \pm 0.1 $ & $ 4.0 \pm 1.3 $ & $ 37 \pm 13 $ & $ 6.7 \pm 2.5 $ & $ 1.5 \pm 1.0 $ & ... & ... & ...\\ 
12 & 05:35:51.57 & -06:34:59.20 & 1 & $ 8700 \pm 700 $ & $ 13.4 \pm 0.3 $ & $ 3.1 \pm 1.0 $ & $ 35 \pm 13 $ & $ 6.6 \pm 2.7 $ & $ 1.3 \pm 0.9 $ & $ 128 \pm 5$ & $ 0.17 \pm 0.07 $ & $ 0.55 \pm 0.03 $\\ 
13 & 05:36:19.45 & -06:29:10.35 & 1 & $ 4800 \pm 500 $ & $ 14.3 \pm 0.0 $ & $ 0.5 \pm 0.2 $ & $ 18 \pm 8 $ & $ 6.2 \pm 3.2 $ & $ 0.3 \pm 0.3 $ & $ 106 \pm 6$ & $ 0.10 \pm 0.06 $ & $ 0.44 \pm 0.03 $\\ 
14 & 05:35:30.01 & -06:26:57.45 & 1 & $ 4500 \pm 600 $ & $ 16.1 \pm 0.1 $ & $ 1.6 \pm 0.5 $ & $ 67 \pm 28 $ & $ 24.9 \pm 12.3 $ & $ 4.7 \pm 3.9 $ & $ 204 \pm 10$ & $ 1.6 \pm 0.8 $ & $ 0.79 \pm 0.04 $\\ 
15 & 05:35:30.78 & -06:26:27.78 & 1 & $ 4900 \pm 600 $ & $ 15.4 \pm 0.1 $ & $ 0.9 \pm 0.3 $ & $ 32 \pm 14 $ & $ 11.1 \pm 5.3 $ & $ 1.1 \pm 0.9 $ & $ 212 \pm 5$ & $ 0.8 \pm 0.4 $ & $ 0.84 \pm 0.02 $\\ 
16 & 05:36:24.88 & -06:24:55.73 & 4 & $ 12700 \pm 600 $ & $ 13.5 \pm 0.3 $ & $ 15.8 \pm 5.1 $ & $ 82 \pm 28 $ & $ 10.8 \pm 3.8 $ & $ 7.1 \pm 4.8 $ & $ 289 \pm 1$ & $ 1.4 \pm 0.5 $ & $ 1.22 \pm 0.02 $\\ 
17 & 05:36:25.33 & -06:22:48.07 & 1 & $ 10600 \pm 600 $ & $ 14.8 \pm 0.5 $ & $ 6.2 \pm 2.0 $ & $ 46 \pm 16 $ & $ 7.4 \pm 2.7 $ & $ 2.3 \pm 1.6 $ & $ 305 \pm 3$ & $ 1.0 \pm 0.4 $ & $ 1.23 \pm 0.03 $\\ 
18 & 05:36:22.11 & -06:23:27.11 & 1 & $ 4100 \pm 500 $ & $ 14.4 \pm 0.1 $ & $ 0.6 \pm 0.2 $ & $ 31 \pm 13 $ & $ 12.9 \pm 6.4 $ & $ 1.0 \pm 0.9 $ & $ 324 \pm 14$ & $ 2.0 \pm 1.0 $ & $ 1.33 \pm 0.06 $\\ 
19 & 05:36:18.99 & -06:22:13.62 & 6 & $ 7600 \pm 700 $ & $ 18.5 \pm 0.5 $ & $ 11.0 \pm 3.5 $ & $ 160 \pm 59 $ & $ 36.1 \pm 15.0 $ & $ 27.9 \pm 20.5 $ & $ 462 \pm 2$ & $ 11.6 \pm 4.8 $ & $ 1.67 \pm 0.02 $\\ 
20 & 05:36:37.44 & -06:15:10.93 & 1 & $ 10000 \pm 1000 $ & $ 14.8 \pm 0.2 $ & $ 1.4 \pm 0.5 $ & $ 12 \pm 5 $ & $ 2.0 \pm 0.9 $ & $ 0.2 \pm 0.1 $ & $ 250 \pm 8$ & $ 0.19 \pm 0.09 $ & $ 1.02 \pm 0.04 $\\ 
21 & 05:35:21.63 & -06:13:13.61 & 1 & $ 7000 \pm 500 $ & $ 15.5 \pm 0.2 $ & $ 0.8 \pm 0.3 $ & $ 14 \pm 6 $ & $ 3.4 \pm 1.5 $ & $ 0.2 \pm 0.2 $ & ... & ... & ...\\ 
22 & 05:35:52.60 & -06:10:05.67 & 1 & $ 6900 \pm 600 $ & $ 15.4 \pm 0.2 $ & $ 1.3 \pm 0.4 $ & $ 23 \pm 9 $ & $ 5.6 \pm 2.4 $ & $ 0.6 \pm 0.4 $ & $ 130 \pm 3$ & $ 0.14 \pm 0.06 $ & $ 0.52 \pm 0.02 $\\ 
23 & 05:36:32.23 & -06:01:14.59 & 1 & $ 7300 \pm 600 $ & $ 15.2 \pm 0.2 $ & $ 2.2 \pm 0.7 $ & $ 35 \pm 13 $ & $ 8.3 \pm 3.5 $ & $ 1.4 \pm 1.0 $ & $ 131 \pm 1$ & $ 0.22 \pm 0.09 $ & $ 0.53 \pm 0.01 $\\ 
24 & 05:35:09.14 & -05:58:10.70 & 1 & $ 12400 \pm 700 $ & $ 15.0 \pm 0.1 $ & $ 4.6 \pm 1.5 $ & $ 25 \pm 9 $ & $ 3.4 \pm 1.3 $ & $ 0.7 \pm 0.5 $ & $ 161 \pm 1$ & $ 0.13 \pm 0.05 $ & $ 0.65 \pm 0.01 $\\ 
25 & 05:35:13.84 & -05:58:03.14 & 1 & $ 6300 \pm 600 $ & $ 15.7 \pm 0.2 $ & $ 2.8 \pm 0.9 $ & $ 59 \pm 22 $ & $ 15.6 \pm 6.5 $ & $ 3.7 \pm 2.7 $ & $ 241 \pm 3$ & $ 1.4 \pm 0.6 $ & $ 0.95 \pm 0.02 $\\ 
26 & 05:35:09.69 & -05:55:44.96 & 4 & $ 12500 \pm 600 $ & $ 15.1 \pm 0.2 $ & $ 10.5 \pm 3.4 $ & $ 56 \pm 19 $ & $ 7.6 \pm 2.7 $ & $ 3.4 \pm 2.3 $ & $ 253 \pm 2$ & $ 0.7 \pm 0.3 $ & $ 1.02 \pm 0.01 $\\ 
27 & 05:35:02.41 & -05:55:41.07 & 1 & $ 12300 \pm 600 $ & $ 15.0 \pm 0.2 $ & $ 5.0 \pm 1.6 $ & $ 27 \pm 9 $ & $ 3.7 \pm 1.3 $ & $ 0.8 \pm 0.5 $ & $ 169 \pm 3$ & $ 0.16 \pm 0.06 $ & $ 0.68 \pm 0.01 $\\ 
28 & 05:33:30.87 & -05:50:40.75 & 1 & $ 6300 \pm 600 $ & $ 15.3 \pm 0.1 $ & $ 0.7 \pm 0.3 $ & $ 15 \pm 6 $ & $ 4.2 \pm 2.0 $ & $ 0.3 \pm 0.2 $ & ... & ... & ...\\ 
29 & 05:34:46.64 & -05:41:59.09 & 2 & $ 6300 \pm 600 $ & $ 16.1 \pm 0.1 $ & $ 0.5 \pm 0.2 $ & $ 11 \pm 5 $ & $ 2.8 \pm 1.4 $ & $ 0.1 \pm 0.1 $ & $ 108 \pm 8$ & $ 0.05 \pm 0.03 $ & $ 0.42 \pm 0.03 $\\ 
30 & 05:34:48.99 & -05:41:38.93 & 1 & $ 4900 \pm 600 $ & $ 16.5 \pm 0.0 $ & $ 0.3 \pm 0.1 $ & $ 9 \pm 5 $ & $ 3.3 \pm 2.0 $ & $ 0.1 \pm 0.1 $ & $ 172 \pm 16$ & $ 0.15 \pm 0.09 $ & $ 0.66 \pm 0.06 $\\ 
31 & 05:34:43.77 & -05:41:26.48 & 1 & $ 6800 \pm 600 $ & $ 16.1 \pm 0.1 $ & $ 1.1 \pm 0.4 $ & $ 19 \pm 7 $ & $ 4.7 \pm 2.1 $ & $ 0.4 \pm 0.3 $ & $ 168 \pm 9$ & $ 0.20 \pm 0.09 $ & $ 0.65 \pm 0.04 $\\ 
32 & 05:34:34.12 & -05:40:02.76 & 2 & $ 9600 \pm 500 $ & $ 16.3 \pm 0.2 $ & $ 1.4 \pm 0.5 $ & $ 12 \pm 5 $ & $ 2.2 \pm 0.8 $ & $ 0.2 \pm 0.1 $ & $ 148 \pm 35$ & $ 0.07 \pm 0.04 $ & $ 0.57 \pm 0.14 $\\ 
33 & 05:34:29.33 & -05:35:47.23 & 1 & $ 7100 \pm 500 $ & $ 15.9 \pm 0.1 $ & $ 0.9 \pm 0.3 $ & $ 15 \pm 6 $ & $ 3.5 \pm 1.5 $ & $ 0.2 \pm 0.2 $ & $ 192 \pm 5$ & $ 0.20 \pm 0.08 $ & $ 0.76 \pm 0.02 $\\ 
34 & 05:35:07.85 & -05:35:48.57 & 1 & $ 8700 \pm 500 $ & $ 19.1 \pm 0.5 $ & $ 5.9 \pm 1.9 $ & $ 66 \pm 22 $ & $ 12.6 \pm 4.6 $ & $ 4.6 \pm 3.1 $ & $ 271 \pm 1$ & $ 1.4 \pm 0.5 $ & $ 0.97 \pm 0.01 $\\ 
35 & 05:35:04.55 & -05:35:00.09 & 1 & $ 9200 \pm 500 $ & $ 20.1 \pm 0.2 $ & $ 5.3 \pm 1.7 $ & $ 53 \pm 18 $ & $ 9.6 \pm 3.4 $ & $ 2.9 \pm 2.0 $ & $ 109 \pm 0$ & $ 0.17 \pm 0.06 $ & $ 0.38 \pm 0.00 $\\ 
36 & 05:35:09.93 & -05:35:04.10 & 1 & $ 5400 \pm 500 $ & $ 19.9 \pm 0.2 $ & $ 1.9 \pm 0.6 $ & $ 56 \pm 21 $ & $ 17.7 \pm 7.8 $ & $ 3.4 \pm 2.6 $ & $ 221 \pm 2$ & $ 1.3 \pm 0.6 $ & $ 0.77 \pm 0.01 $\\ 
37 & 05:34:40.76 & -05:31:07.24 & 1 & $ 15900 \pm 600 $ & $ 17.9 \pm 0.3 $ & $ 4.4 \pm 1.4 $ & $ 14 \pm 5 $ & $ 1.5 \pm 0.5 $ & $ 0.2 \pm 0.2 $ & $ 233 \pm 3$ & $ 0.13 \pm 0.04 $ & $ 0.86 \pm 0.02 $\\ 
38 & 05:33:57.28 & -05:23:33.04 & 1 & $ 8600 \pm 700 $ & $ 19.4 \pm 0.6 $ & $ 2.4 \pm 0.8 $ & $ 28 \pm 10 $ & $ 5.4 \pm 2.2 $ & $ 0.8 \pm 0.6 $ & $ 180 \pm 2$ & $ 0.3 \pm 0.1 $ & $ 0.64 \pm 0.01 $\\ 
39 & 05:35:19.64 & -05:15:25.82 & 6 & $ 5800 \pm 600 $ & $ 25.2 \pm 0.7 $ & $ 2.0 \pm 0.7 $ & $ 51 \pm 20 $ & $ 15.0 \pm 6.5 $ & $ 2.8 \pm 2.1 $ & $ 354 \pm 2$ & $ 2.8 \pm 1.2 $ & $ 1.10 \pm 0.02 $\\ 
40 & 05:35:26.82 & -05:09:58.50 & 6 & $ 13900 \pm 600 $ & $ 21.9 \pm 1.0 $ & $ 35.7 \pm 11.7 $ & $ 155 \pm 53 $ & $ 18.9 \pm 6.7 $ & $ 25.8 \pm 17.6 $ & $ 405 \pm 1$ & $ 4.7 \pm 1.7 $ & $ 1.35 \pm 0.03 $\\ 
41 & 05:35:24.45 & -05:08:27.86 & 1 & $ 4600 \pm 600 $ & $ 18.7 \pm 0.2 $ & $ 3.0 \pm 0.9 $ & $ 117 \pm 48 $ & $ 41.8 \pm 20.4 $ & $ 14.1 \pm 11.4 $ & $ 302 \pm 2$ & $ 5.7 \pm 2.8 $ & $ 1.09 \pm 0.01 $\\ 
42 & 05:35:24.23 & -05:07:52.70 & 2 & $ 3800 \pm 600 $ & $ 18.5 \pm 0.2 $ & $ 2.4 \pm 0.8 $ & $ 140 \pm 63 $ & $ 62.2 \pm 35.6 $ & $ 21.0 \pm 18.9 $ & $ 375 \pm 3$ & $ 13.2 \pm 7.5 $ & $ 1.36 \pm 0.01 $\\ 
43 & 05:35:27.86 & -05:07:13.42 & 1 & $ 5100 \pm 500 $ & $ 19.5 \pm 0.1 $ & $ 1.2 \pm 0.4 $ & $ 38 \pm 15 $ & $ 12.9 \pm 5.7 $ & $ 1.6 \pm 1.2 $ & $ 281 \pm 6$ & $ 1.5 \pm 0.7 $ & $ 0.99 \pm 0.02 $\\ 
44 & 05:35:31.71 & -05:05:54.01 & 3 & $ 9400 \pm 600 $ & $ 19.1 \pm 0.1 $ & $ 5.0 \pm 1.6 $ & $ 48 \pm 17 $ & $ 8.6 \pm 3.2 $ & $ 2.5 \pm 1.7 $ & $ 314 \pm 5$ & $ 1.3 \pm 0.5 $ & $ 1.12 \pm 0.02 $\\ 
45 & 05:35:26.04 & -05:05:42.70 & 4 & $ 3200 \pm 600 $ & $ 17.7 \pm 0.2 $ & $ 1.9 \pm 0.6 $ & $ 160 \pm 79 $ & $ 80.9 \pm 52.2 $ & $ 25.9 \pm 25.5 $ & $ 257 \pm 2$ & $ 8.0 \pm 5.2 $ & $ 0.95 \pm 0.01 $\\ 
46 & 05:35:19.69 & -05:05:03.91 & 1 & $ 12500 \pm 600 $ & $ 19.7 \pm 0.5 $ & $ 7.1 \pm 2.3 $ & $ 38 \pm 13 $ & $ 5.2 \pm 1.8 $ & $ 1.6 \pm 1.1 $ & $ 649 \pm 22$ & $ 3.3 \pm 1.2 $ & $ 2.28 \pm 0.09 $\\ 
47 & 05:35:26.35 & -05:03:42.54 & 2 & $ 9400 \pm 700 $ & $ 18.3 \pm 0.7 $ & $ 6.9 \pm 2.3 $ & $ 66 \pm 24 $ & $ 12.0 \pm 4.7 $ & $ 4.7 \pm 3.4 $ & $ 204 \pm 2$ & $ 0.8 \pm 0.3 $ & $ 0.74 \pm 0.02 $\\ 
48 & 05:35:23.08 & -05:01:25.07 & 3 & $ 5800 \pm 600 $ & $ 17.3 \pm 0.2 $ & $ 12.3 \pm 3.9 $ & $ 307 \pm 117 $ & $ 87.3 \pm 38.0 $ & $ 97.8 \pm 73.5 $ & $ 299 \pm 2$ & $ 11.8 \pm 5.1 $ & $ 1.12 \pm 0.01 $\\ 
49 & 05:35:20.12 & -05:00:50.24 & 1 & $ 5100 \pm 600 $ & $ 16.7 \pm 0.2 $ & $ 6.2 \pm 2.0 $ & $ 200 \pm 80 $ & $ 64.9 \pm 32.0 $ & $ 41.7 \pm 33.8 $ & $ 216 \pm 1$ & $ 4.6 \pm 2.2 $ & $ 0.82 \pm 0.01 $\\ 
50 & 05:35:18.15 & -05:00:24.76 & 3 & $ 4200 \pm 600 $ & $ 17.4 \pm 0.1 $ & $ 4.5 \pm 1.5 $ & $ 216 \pm 93 $ & $ 85.6 \pm 45.4 $ & $ 49.0 \pm 41.8 $ & $ 302 \pm 2$ & $ 11.7 \pm 6.2 $ & $ 1.13 \pm 0.01 $\\ 
51 & 05:35:34.20 & -04:59:49.22 & 1 & $ 5900 \pm 600 $ & $ 19.3 \pm 0.1 $ & $ 0.6 \pm 0.2 $ & $ 14 \pm 6 $ & $ 4.1 \pm 1.8 $ & $ 0.2 \pm 0.2 $ & $ 223 \pm 6$ & $ 0.3 \pm 0.1 $ & $ 0.79 \pm 0.02 $\\ 
52 & 05:35:29.73 & -04:59:43.75 & 1 & $ 3400 \pm 500 $ & $ 18.1 \pm 0.1 $ & $ 0.8 \pm 0.2 $ & $ 55 \pm 24 $ & $ 27.9 \pm 16.0 $ & $ 3.3 \pm 3.0 $ & $ 189 \pm 9$ & $ 1.5 \pm 0.9 $ & $ 0.69 \pm 0.03 $\\ 
53 & 05:35:29.77 & -04:58:46.11 & 1 & $ 4400 \pm 600 $ & $ 18.1 \pm 0.2 $ & $ 2.4 \pm 0.8 $ & $ 102 \pm 43 $ & $ 39.5 \pm 20.0 $ & $ 11.3 \pm 9.4 $ & $ 660 \pm 11$ & $ 25.8 \pm 13.1 $ & $ 2.41 \pm 0.04 $\\ 
\enddata
\tablecomments{(1) The index of the cores. (2) and (3) The coordinate of the center of the cores. (4) $N_{\mathrm{yso}}$ is the number of protostars. (5) The radius of the dense core is defined to be the geometric mean of 2$\sigma$ widths of the semimajor and semiminor axes of the dense core. (6) $T_d$ is the dust temperature. (7) The core mass is estimated from the JCMT 850 $\mu$m continuum emission. (8) N(H$_2$) is the column density. (9) $\rho$ is the volumn density. (10) $u_G$ is the gravitational energy density. (11) $\sigma_{\mathrm{NT}}$ is the 1$\sigma$ non-thermal linewidth. (12) $P_T$ is the turbulent pressure. (13) $\mathcal{M}$ is the Mach number.}

\end{deluxetable*}
\label{tab:core}
\end{longrotatetable}

\begin{table}[h]
    \centering
    \begin{tabular}{ccc}
        \hline
        index & $B_{pos}$ (mG) & $\lambda$ \\
        \hline
        13 & $0.04^{+0.04}_{-0.02}$ & $3.1^{+4.3}_{-2.0}$ \\ 
        16 & $0.4^{+0.2}_{-0.1}$ & $1.6^{+1.1}_{-0.7}$ \\ 
        17 & $0.3^{+0.2}_{-0.1}$ & $1.1^{+0.8}_{-0.6}$ \\ 
        19 & $2.2^{+1.8}_{-0.9}$ & $0.5^{+0.5}_{-0.3}$ \\ 
        24 & $0.08^{+0.08}_{-0.04}$ & $2.3^{+2.6}_{-1.4}$ \\ 
        25 & $0.2^{+0.2}_{-0.1}$ & $2.2^{+2.9}_{-1.3}$ \\ 
        26 & $0.16^{+0.07}_{-0.05}$ & $2.7^{+1.8}_{-1.2}$ \\ 
        27 & $0.09^{+0.05}_{-0.03}$ & $2.3^{+1.5}_{-1.1}$ \\ 
        31 & $0.12^{+0.12}_{-0.05}$ & $1.1^{+1.2}_{-0.7}$ \\ 
        34 & $0.8^{+0.4}_{-0.2}$ & $0.6^{+0.4}_{-0.3}$ \\ 
        35 & $0.09^{+0.05}_{-0.03}$ & $4.1^{+2.8}_{-1.9}$ \\ 
        36 & $0.8^{+0.8}_{-0.4}$ & $0.5^{+0.6}_{-0.3}$ \\ 
        37 & $0.06^{+0.05}_{-0.02}$ & $1.7^{+1.5}_{-0.9}$  \\ 
        39 & $0.6^{+0.4}_{-0.2}$ & $0.7^{+0.6}_{-0.3}$ \\ 
        40 & $0.30^{+0.07}_{-0.07}$ & $4.0^{+1.9}_{-1.5}$ \\ 
        41 & $0.7^{+0.7}_{-0.3}$ & $1.2^{+1.5}_{-0.7}$ \\ 
        42 & $0.8^{+0.6}_{-0.4}$ & $1.3^{+1.7}_{-0.7}$ \\ 
        43 & $0.2^{+0.2}_{-0.1}$ & $1.3^{+1.9}_{-0.8}$ \\ 
        44 & $0.3^{+0.2}_{-0.1}$ & $1.2^{+0.9}_{-0.6}$ \\ 
        46 & $0.7^{+0.3}_{-0.2}$ & $0.4^{+0.2}_{-0.2}$ \\ 
        47 & $0.5^{+0.3}_{-0.2}$ & $1.1^{+0.8}_{-0.6}$ \\ 
        48 & $1.8^{+0.7}_{-0.6}$ & $1.3^{+0.9}_{-0.6}$ \\ 
        49 & $2.2^{+1.2}_{-0.9}$ & $0.7^{+0.6}_{-0.3}$ \\ 
        50 & $2.7^{+1.6}_{-1.2}$ & $0.6^{+0.6}_{-0.3}$ \\ 
        51 & $0.12^{+0.15}_{-0.06}$ & $0.8^{+1.2}_{-0.5}$ \\ 
        52 & $0.3^{+0.4}_{-0.2}$ & $1.2^{+1.9}_{-0.8}$ \\ 
        53 & $3.0^{+2.6}_{-1.4}$ & $0.3^{+0.3}_{-0.1}$ \\ 
        \hline
    \end{tabular}
    \caption{Magnetic field strength and mass-to-flux ratio of the dense cores.}
    \label{tab:B}
\end{table}

\section{Results} \label{sec:results}
To investigate the relationship between dense core properties and fragmentation, we compared the cumulative distribution functions (CDF) of the physical parameters of the two groups, the dense cores with single and binary/multiple systems.
To estimate the uncertainties of the CDF, 
we randomly varied each measurement assuming a Gaussian probability distribution with the center as the original measured value and the 1$\sigma$ as the measurement uncertainty from error propagation, and then calculated a CDF from the varied measurements.
We repeated this process 10000 times, generated a sample of CDF, and adopted the 50th percentile at each step as the final CDF and the 16th and 84th percentiles at each step as the 1$\sigma$ error of the CDF.

\begin{figure*}
    \centering  
    \subfigure{
        \includegraphics[width=0.45\textwidth]{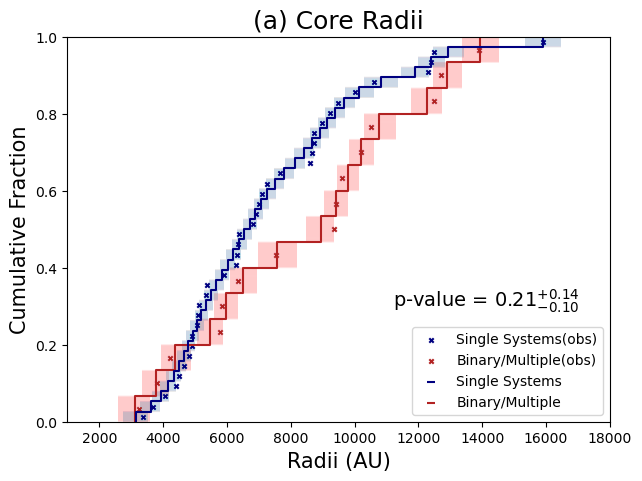}\label{fig:radii}
    }
    \subfigure{
        \includegraphics[width=0.45\textwidth]{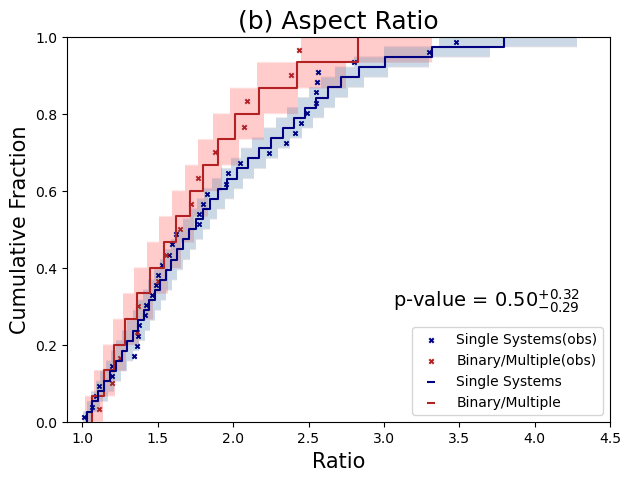}\label{fig:ab}
    }
    
    \subfigure{
        \includegraphics[width=0.45\textwidth]{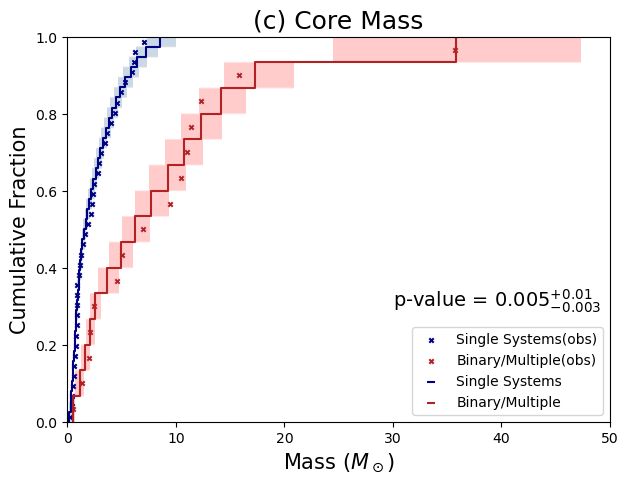}\label{fig:mass}
    }
    \subfigure{
        \includegraphics[width=0.45\textwidth]{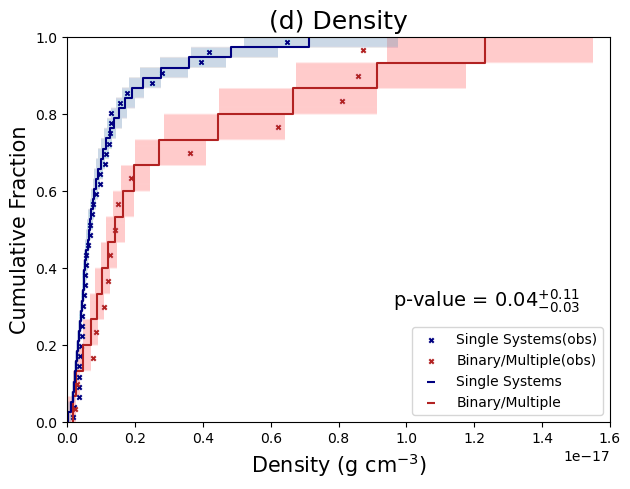}\label{fig:density}   
    }
    \caption{Cumulative distributions (solid lines)
    of the (a) radius, (b) aspect ratio, (c) mass, and (d) density of the dense cores with single (red) and binary/multiple (blue) systems in our sample. 
    The light-blue and pink shaded regions represent the 1$\sigma$ errors of the cumulative distributions. The data points denote the original cumulative distributions directly extracted from the measurements before accounting for the measurement uncertainties.}
    \label{fig:radii_mass}
\end{figure*}

\subsection{Core Morphology}\label{sec:size and morphology}
Figure \ref{fig:radii} and \ref{fig:ab} compare the CDF of the radii and aspect ratio of the dense cores with single and binary/multiple systems.
Although the dense cores with multiple systems tend to have larger radii, the overall range of the two groups is similar.
The CDF of the aspect ratio of the two groups are also similar, approximately consistent within the uncertainties. 
These show that the fragmented dense cores do not have particular size or elongation. 
We adopted the Kolmogorov–Smirnov (KS) test for statistical comparison between the CDF of the two groups.
The p-values from the KS test are $0.21^{+0.14}_{-0.10}$ and $0.50^{+0.32}_{-0.29}$, suggesting that there is indeed no significant difference between the core radii and aspect ratio of the dense cores with single and binary/multiple systems.

\subsection{Density and Thermal Jeans Instability}\label{sec:density and thermal}

\begin{figure}[h]
    \centering
    \includegraphics[width=0.95\linewidth]{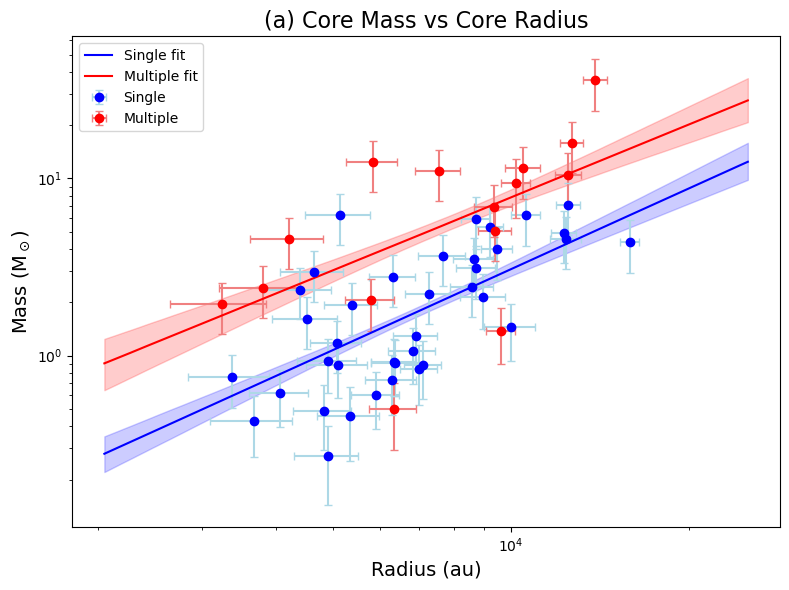}\label{fig:core_mass_radii}
    \includegraphics[width=0.97\linewidth]{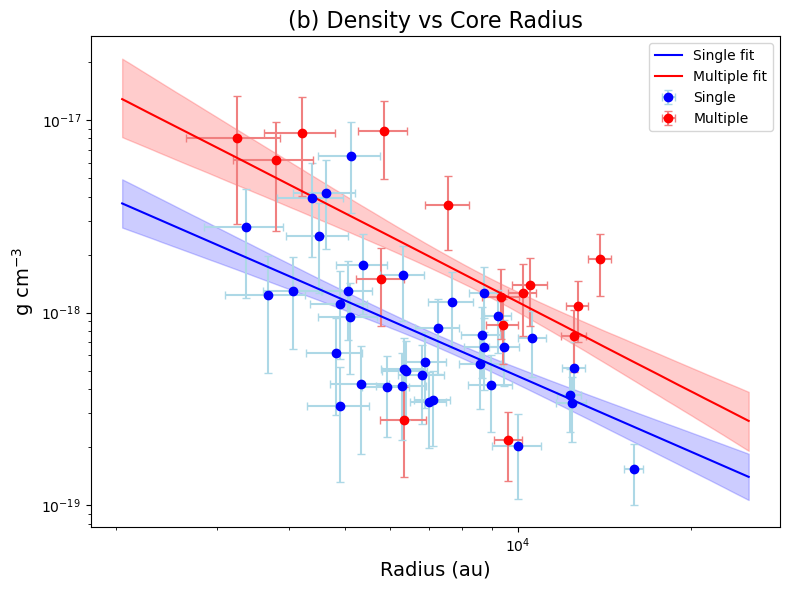}\label{fig:core_density_radii}
    \caption{(a) core mass and (b) density versus core radii. The blue dots show the cores with single protostar. The red dots show the cores with binary/multiple system. Solid lines show the best-fit power-law relations derived via bootstrapping.
    Shaded regions represent the 1$\sigma$ confidence intervals of the fits.}
    \label{fig:core_mass_dens_radii}
\end{figure}
\par
Figure \ref{fig:mass} compares the core mass of single and binary/multiple systems.
The dense cores forming binary/multiple systems tend to be more massive than those forming single systems.
The p-value from the KS test on their CDF is $0.005^{+0.01}_{-0.003}$. 
Since the dense cores with single and binary/multiple systems have similar radii (Figure \ref{fig:radii}), the dense cores with binary/multiple systems exhibit significantly higher densities.
The KS test on the CDF of the density of the dense cores with single and binary/multiple systems results in a p-value of $0.04^{+0.11}_{-0.03}$.
This also can been seen in the mass--radius relations of these dense cores (Figure \ref{fig:core_mass_dens_radii})
For a given core size, the dense cores forming binary/multiple systems tend to have higher mass and thus density than those forming single systems. 
The power-law fitting to the mass--radius and density--radius relations suggests that the mass and density tend to be a factor of 3 higher in the dense cores forming binary/multiple systems.

\par
We note that a part of the initial core mass has been converted to its protostellar mass, which is not included here (Section \ref{sec:density}), so the total enclosed mass in the cores can be underestimated. 
Nevertheless, the total protostellar mass in the cores is expected to correlate with the number of the protostars, and thus these observed tendencies are unlikely to be affected by the protostellar mass.

\begin{figure*}
    \centering
    \subfigure{
        \includegraphics[width=0.4\textwidth]{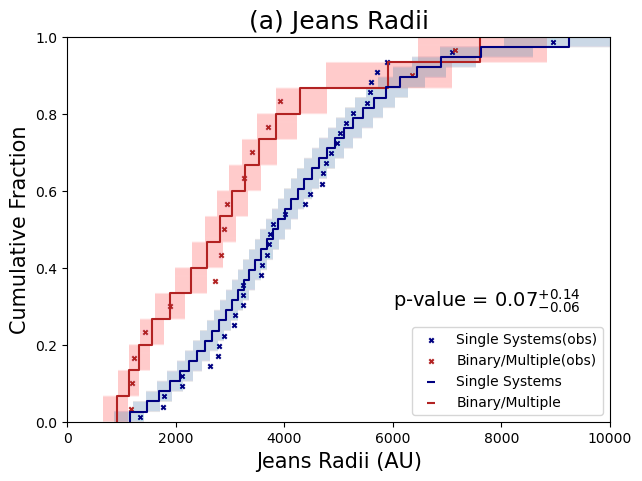}\label{fig:jeansradii}}
    \subfigure{
        \includegraphics[width=0.4\textwidth]{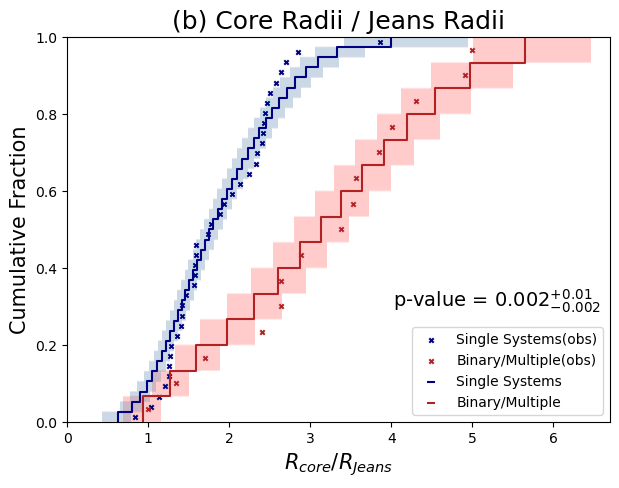}\label{fig:jeansratio}}

    \subfigure{
        \includegraphics[width=0.4\textwidth]{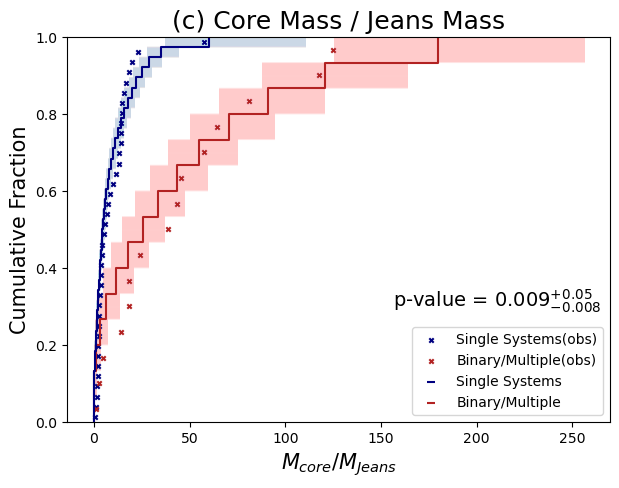}\label{fig:jeansmassratio}}
    \subfigure{
    \label{fig:mach}
    \includegraphics[width=0.4\textwidth]{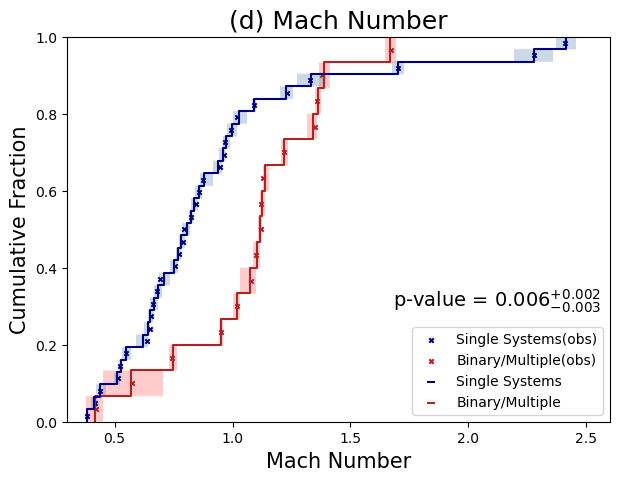}}

    \subfigure{
    \label{fig:PtuG}
    \includegraphics[width=0.4\textwidth]{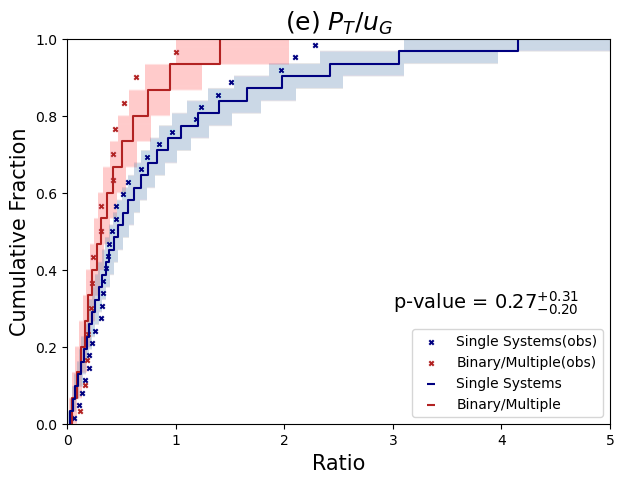}}
    \subfigure{
    \label{fig:B}
    \includegraphics[width=0.4\textwidth]{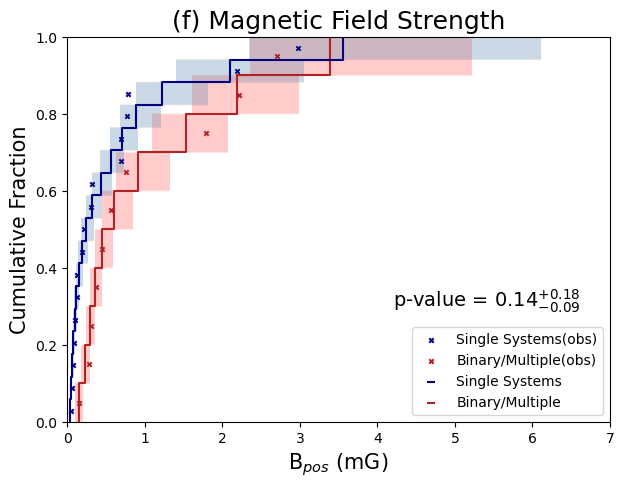}}

    \subfigure{
    \label{fig:lambda}
    \includegraphics[width=0.4\textwidth]{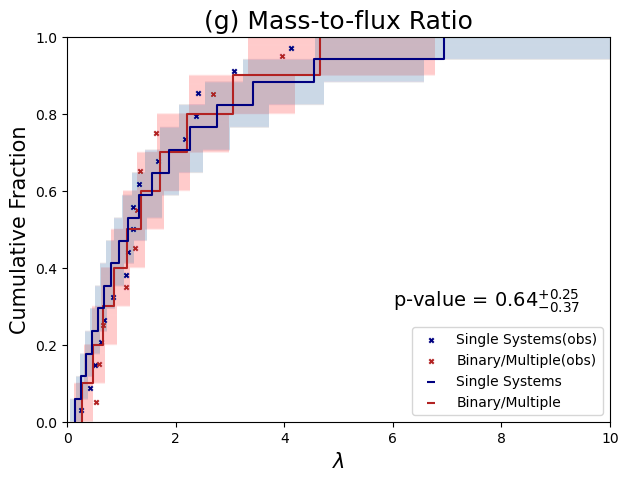}}
    
    \caption{
    Same as Fig.~3 but for 
    (a) Jeans radius, (b) core radius over Jeans radius, (c) core mass over Jeans mass, (d) Mach number, (e) turbulent over gravitational energy, (f) magnetic field strength, and (g) mass-to-flux ratio of the dense cores.
    }
    \label{fig:thermal_turb_B}
\end{figure*}

\par
Figure \ref{fig:jeansradii}, \ref{fig:jeansratio}, and \ref{fig:jeansmassratio}compare the CDF of the Jeans radii, the ratios of the core radii to Jeans radii, and the ratios of the core mass to Jeans mass.
Although the dense cores with single systems have a larger median Jeans radius than those with binary/multiple systems, the ranges of the Jeans radii of the two groups are similar. 
There is no significant difference between their Jeans radii, and the p-value from the KS test on their CDF is $0.07^{+0.14}_{-0.06}$. 
However, as shown in Figure \ref{fig:core_mass_dens_radii}, for a given size, dense cores forming binary/multiple systems tend to have higher density and thus smaller Jean radii and mass (Eq.~\ref{eq:Jeans_length}).
After taking the ratios of the core radii and mass to Jeans radii and mass, the cores with binary/multiple systems clearly show larger ratios with a p-value of $0.002^{+0.01}_{-0.002}$ and $0.009^{+0.05}_{-0.008}$ from the KS test, respectively.

\subsection{Turbulent Velocity and Energy}\label{sec:tub}
\par
Figure \ref{fig:mach} compares the Mach numbers in the dense cores with single and binary/multiple systems.
The Mach numbers in the dense cores with binary/multiple systems are significantly higher than those with single systems with a p-value of $0.006^{+0.002}_{-0.003}$ from the KS test. 
In the 31 single systems, excluding those without N$_2$H$^+$ (1--0) data, 24 of them (77\%) have subsonic and 7 of them (23\%) have supersonic turbulence.
In contrast, in the 15 binary/multiple systems, 4 of them (27\%) have subsonic and 11 of them (73\%) have supersonic turbulence.

\par
Figure \ref{fig:PtuG} presents the CDF of energy ratios of $P_T/u_G$ of the two groups.
The dense cores with single systems tend to have larger ratios of the turbulent to gravitational energy, and the maximum ratio in the sample is observed in the single systems. 
This may hint that the substantial contribution by the turbulence compared to the gravity can suppress the core fragmentation.
Nevertheless, the ratios between the two groups are not statistically significantly different with a p-value of $0.27^{+0.31}_{-0.20}$ from the KS test.

All the binary/multiple systems and approximately 80\% of the single systems have this ratio smaller than 1, showing that the gravity is more dominant as expected. 
We note that a few sources have this ratio larger than 1. This could be due to the possible underestimation of the the gravitational energy density because the protostellar mass was not included in the calculation. 

\subsection{Magnetic field and mass-to-flux ratio}\label{sec:B_ab}

\par
Figure \ref{fig:B} shows that the dense cores with binary/multiple systems tend to have larger magnetic field strengths than those with single systems.
The KS test result shows a p-value of $0.14^{+0.18}_{-0.09}$. Thus, this difference is likely not significant.
When comparing the mass-to-flux ratio (Figure \ref{fig:lambda}), the CDF of the dense cores with single and binary/multiple systems are almost identical and overlap within the uncertainties. The KS test results in a p-value of $0.64^{+0.25}_{-0.37}$. Therefore, there is no difference in mass-to-flux ratios between the single and binary/multiple systems in Orion A.

\par
We note that there are 9 dense cores with a mass-to-flux ratio smaller than 1, although they are collapsing to form stars and are expected to be gravity-dominated. 
This could be due to the underestimation of the enclosed mass because the protostellar mass is unknown and not included, or the gravity is only locally dominated in the dense cores.

\begin{figure}
    \centering
    \subfigure{
    \label{fig:abRatio}
    \includegraphics[width=0.45\textwidth]{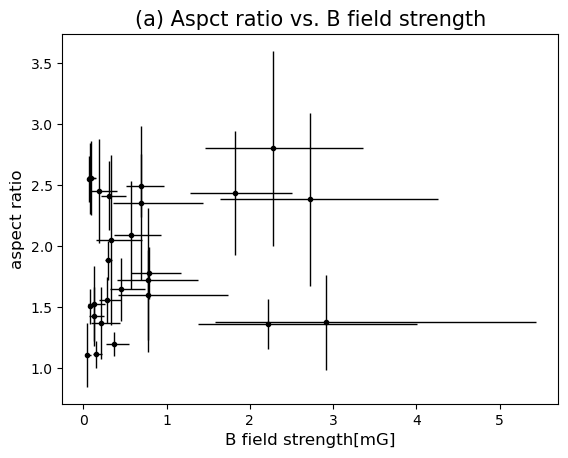}}
    \subfigure{
    \label{fig:deltatheta_b}
    \includegraphics[width=0.45\textwidth]{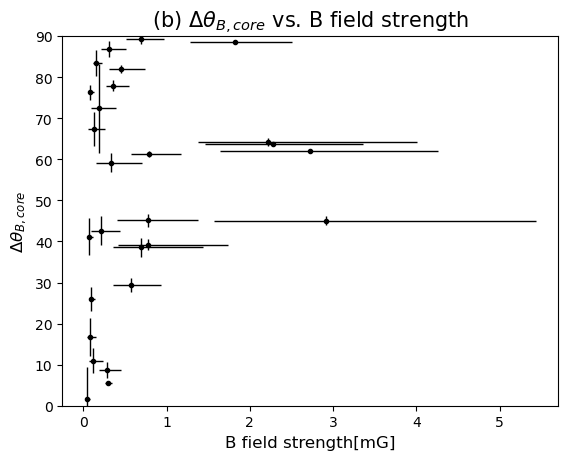}}
    \caption{(a) Ratio of the major to minor axes of the dense cores (aspect ratio) versus their magnetic field strengths.
    (b) Angle between the magnetic field orientation and major axis of the dense cores ($\Delta\theta_{B,core}$) versus the magnetic field strength.}
\end{figure}

Figure \ref{fig:abRatio} and \ref{fig:deltatheta_b} compare the aspect ratio and orientation of the dense cores with the magnetic field properties in Orion~A. 
The aspect ratios of the dense cores do not correlate with the magnetic field strengths with a p-value of $0.7$ from Spearman correlation. 
Highly elongated dense cores with weak magnetic fields are present. 
The orientations of the dense cores with respect to the magnetic fields also do not correlated with the magnetic field strengths with a p-value of $0.1$ from Spearman correlation. 
The dense cores in Orion~A do not prefer to be parallel or perpendicular to the magnetic field orientation regardless of the magnetic field strength.

\begin{figure*}
    \centering
    \subfigure{
        \includegraphics[width=0.45\textwidth]{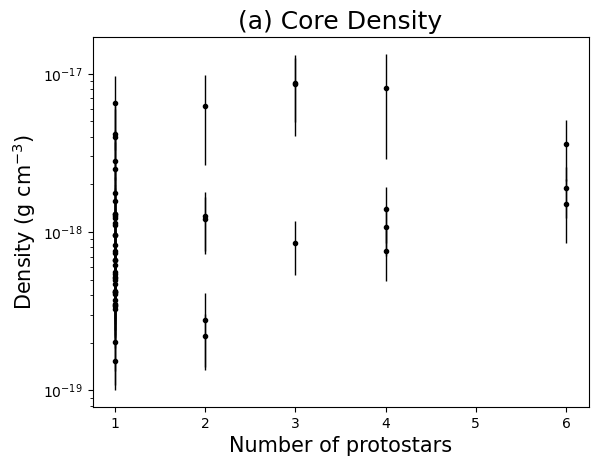}\label{fig:density_number}
    }
    \quad
    \subfigure{
        \includegraphics[width=0.45\textwidth]{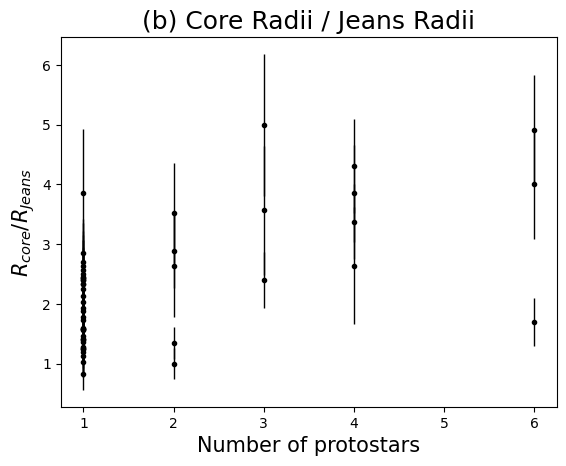}\label{fig:coreRadiijeansRadii_number}   
    }
    \quad
    \subfigure{
        \includegraphics[width=0.45\textwidth]{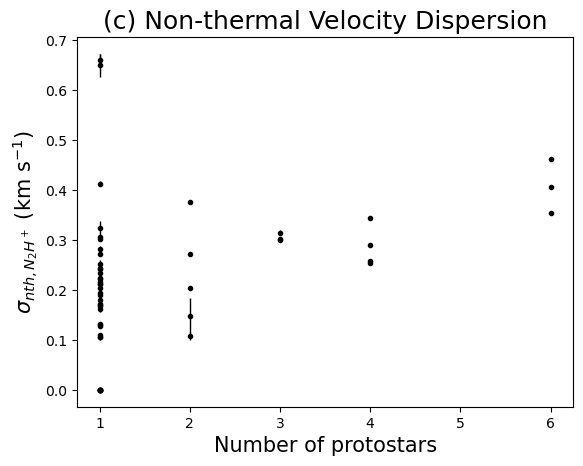}\label{fig:nthvd_number}   
    }
    \quad
    \subfigure{
        \includegraphics[width=0.45\textwidth]{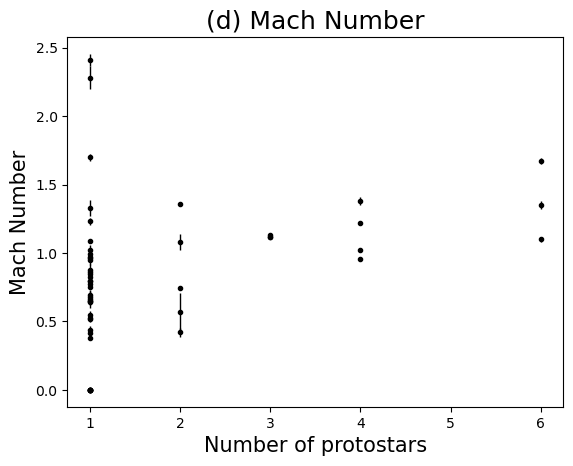}\label{fig:mach_number}   
    }
    \caption{Numbers of protostars in the dense cores versus (a) mass density, (b) ratio of core radius to Jeans radius, (c) non-thermal velocity dispersion, and (d) Mach number of the dense cores.}
    \label{fig:prop_number}
\end{figure*}

\subsection{Physical conditions and number of fragments}
For the parameters showing significant difference between the single and binary/multiple systems, we also compared them with the number of protostars in the dense cores and calculated Spearman correlation (Figure \ref{fig:prop_number}).
The number of protostars in the dense cores is correlated with the core mass, density, the ratios of core radii to Jeans radii and core mass to Jeans mass, the non-thermal line width, and the Mach number.
The p-values from Spearman correlations of the number of protostars and these parameters are all 0.01 or smaller. 

\section{Discussion} \label{sec:discussion}
\subsection{Dense core properties and fragmentaion}\label{sec:dis_properties}
We compared the physical conditions of the dense cores with single systems and binary/multiple systems, which is summarized in Table \ref{tab:compare}.
We found that statistically the dense cores with binary/multiple systems have higher core mass, density, Mach number, ratios of core radii to Jeans radii, and ratios of core mass to Jeans mass than those with single systems (Figure \ref{fig:radii_mass}, \ref{fig:core_mass_dens_radii}, and \ref{fig:thermal_turb_B}).
In addition, the Mach numbers in the dense cores with binary/multiple systems are mostly larger than 1. 
While there is no statistically significant difference in the core size, morphology, energy ratio of turbulence to gravity, and mass-to-flux ratio between the dense cores with single and binary/multiple systems (Figure \ref{fig:radii_mass} and \ref{fig:thermal_turb_B}).
Therefore, our results suggest that fragmentation is prone to occur in high-density dense cores with supersonic turbulent velocities, where the size and mass are larger than their Jeans length and mass.

\begin{table*}[]
    \centering
    \begin{tabular}{ccc}
        \hline
        Parameters & Results & p-value from KS Test\\
        \hline
        Core radii & No difference & $0.21^{+0.14}_{-0.10}$\\
        Aspect Ratio & No difference & $0.50^{+0.32}_{-0.29}$\\
        Core mass & Binary/Multiple $>$ Single & $0.005^{+0.01}_{-0.003}$\\
        Density & Binary/Multiple $>$ Single & $0.04^{+0.11}_{-0.03}$\\
        Jeans radii & No difference & $0.07^{+0.14}_{-0.06}$\\
        Core radii/Jeans radii & Binary/Multiple $>$ Single & $0.002^{+0.01}_{-0.002}$\\
        Core mass/Jeans mass & Binary/Multiple $>$ Single & $0.009^{+0.05}_{-0.008}$\\
        Mach number & Binary/Multiple $>$ Single & $0.006^{+0.002}_{-0.003}$\\
        $P_T/u_G$ & No difference & $0.27^{+0.31}_{-0.20}$\\
        Magnetic field strength & No difference & $0.14^{+0.18}_{-0.09}$\\
        Mass-to-flux ratio & No difference & $0.64^{+0.25}_{-0.37}$\\
        \hline
    \end{tabular}
    \caption{Summary of the KS test results of core properties between single and binary/multiple systems}.
    \label{tab:compare}
\end{table*}
\par
Fragmentation and properties of dense cores in the Orion molecular cloud complex have also been studied by \cite{2022ApJ...931..158L}. 
They identified 29 dense cores with single system and 14 dense core cores with binary/multiple systems based on the number of compact sources detected in the ALMA 1.3 mm continuum maps at an angular resolution 0\farcs35 \citep{2020ApJS..251...20D}. 
Nine of their sample sources are also included in our sample. 
Among them, G208.68-19.20N1 was identified as a single system in \cite{2022ApJ...931..158L}, while there were 3 protostars in the core detected with the VANDAM survey \citep{2020ApJ...890..130T}, so it was classified as a multiple system in our sample.
\cite{2022ApJ...931..158L} also found that the gas density and Mach number in the dense cores with binary/multiple systems are higher than those with single systems, although their comparison of core mass did not show a significant difference between single and binary/multiple systems.
Thus, the two different samples of 91 dense cores in total, both suggest that the gas density and Mach number are likely the key factors in the core fragmentation in the Orion molecular cloud complex.

\begin{figure}
    \centering
    \includegraphics[width=0.95\linewidth]{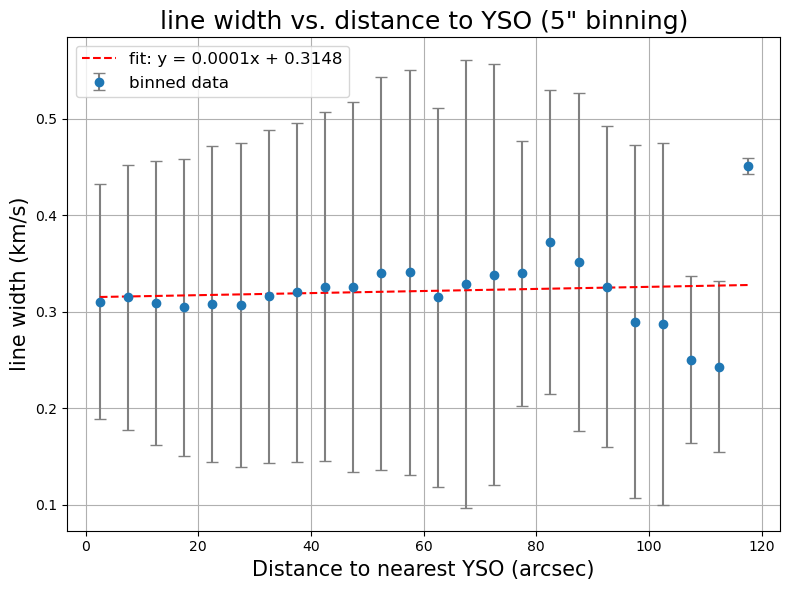}
    \caption{Mean non-thermal line width as a function of projected angular distance to the nearest protostar, measured in 5$\arcsec$ bins. Error bars show the standard deviation within each bin. The red dashed line shows the best-fit linear trend.}
    \label{fig:lw_yso}
\end{figure}

\par
We note that the high Mach number in the dense cores with binary and multiple systems in our sample is unlikely caused by the energy input from star formation activities, such as protostellar outflows or infalling motion. 
If the non-thermal linewidth is affected by outflows or infalling motion, we expected to observe higher velocity dispersions near the protostars, where the outflows and gravitational collapse are more prominent.
We have measured the non-thermal velocity dispersion ($\sigma_{NT}$) as a function of the distance to the nearest protostars, and found that the turbulent velocity does not correlate with the distance to the protostars (Figure \ref{fig:lw_yso}).
Furthermore, N$_2$H$^+$ traces the cold and dense gas, and is not sensitive to the warm gas around outflow cavities or inner protostellar envelopes \citep{1992ApJ...387..417W}.

\subsection{Turbulent Core Fragmentation}
These results are consistent with the expectation from the turbulent core fragmentation scenario \citep{2004A&A...414..633G, 2007prpl.conf..133G,2010ApJ...725.1485O}.
Dense cores experiencing supersonic turbulence are prone to forming multiple high density peaks \citep{2004A&A...414..633G, 2007prpl.conf..133G}. 
If the overall density of the dense core is high, these local density peaks are more likely to exceed the local Jeans mass, and the dense core fragments \citep{2004A&A...414..633G, 2007prpl.conf..133G,2010ApJ...725.1485O}.
Indeed, our results show that the dense cores with binary/multiple systems in Orion~A tend to have higher ratios of core radii to Jeans radii and higher ratios of core mass to Jeans mass than those with single systems. 
Furthermore, the numbers of the protostars in the dense cores in our sample tend to correlate with their mass density and ratios of the core radii to Jeans radii with p-values from Spearman correlation as $0.01^{+0.06}_{-0.01}$ and $0.0008^{+0.004}_{-0.0007}$, respectively (Figure \ref{fig:prop_number}). 
These correlations have also been observed in more massive dense cores of ten to hundreds of solar masses \citep{2015MNRAS.453.3785P, 2021ApJ...912..159P}.
Besides, the observations comparing the mass distribution of dense cores in clumps and that of stars in stellar clusters suggest that massive dense cores tend to fragment and form more stars \citep{2024ApJS..270....9X}.
Thus, this tendency that fragmentation is prone to occur in high-density cores with larger ratio of their radii to Jeans radii is likely valid in both massive and low-mass dense cores, and multiplicity tends to increase with ratio of core sizes to Jeans lengths or core mass to Jeans mass. 

In addition, in our sample, we found that the numbers of the protostars also tend to correlate with the non-thermal velocity dispersion and Mach number in the dense cores with p-values from Spearman correlation as $0.0005^{+0.0002}_{-0.0001}$ and $0.0007^{+0.0003}_{-0.0002}$, respectively (Figure \ref{fig:prop_number}).
This supports the scenario that higher turbulent velocity promotes the formation of multiple density peaks in dense cores. 
Nevertheless, this correlation is not observed in the other sample of dense cores with even stronger turbulence, where the non-thermal line widths measured in the N$_2$H$^+$ (1--0) line are larger than 0.6 km s$^{-1}$\citep{2015MNRAS.453.3785P}, significantly larger than those in our sample. 
Thus, the density can be a more general factor in controlling subsequent local collapse and influencing multiplicity compared to the turbulent velocity in dense cores.

\par
Besides inducing density peaks, turbulence may provide support against gravitational collapse and suppress multiplicity \citep{2007prpl.conf...63B, 2012A&ARv..20...55H, 2024MNRAS.528.1460R}.
Although in our sample the turbulence tends to be stronger in the dense cores with binary/multiple systems, their ratios of the turbulent pressure to gravitational energy density are comparable to those in the dense cores with single systems (Figure \ref{fig:PtuG}). 
Thus, the turbulent support against gravity does not affect the occurrence of fragmentation in the dense cores in Orion~A. 
This is likely because in our sample the gravity is mostly dominant over turbulence in the dense cores with binary/multiple systems with the ratios of the turbulent pressure to gravitational energy density smaller than 1. 

\par 
Nevertheless, in more turbulent environments, such as the infrared dark cloud G11.11$-$0.12 with an average Mach number of 3, it is observed that turbulent support could affect the mass and separation of the fragments \citep{2014MNRAS.439.3275W}, although it is not seen in other turbulent massive dense cores with similar Mach numbers \citep{2015MNRAS.453.3785P}. 
Our data are not able to probe the mass of fragments and the physical conditions in their local environments inside the dense cores in Orion~A, so we cannot investigate whether the turbulent support affects the mass and separation of fragments. 
Further studies on spatially resolved physical conditions and dynamics in these dense cores are needed to further investigate their fragmentation process.

\subsection{The Role of Magnetic Field in Fragmentation}\label{sec:Binfragmentation}
\par
Our results show that the dense cores with binary/multiple systems may have stronger magnetic fields (Figure \ref{fig:B}). However, it is not statistically significant.
The magnetic field strength in star-forming regions is known to increase with density \citep{2012ARA&A..50...29C}.
Therefore, the possible difference in the magnetic field strengths in the dense core with single and binary/multiple systems may be due to their difference in density.
When comparing the magnetic field strength to gravity,
the mass-to-flux ratios in the dense cores with single and binary/multiple systems are almost identical (Figure \ref{fig:lambda}). 
Therefore, there is no sign that the magnetic support affects the occurrence of fragmentation in the dense cores in Orion~A.

\par
In addition, the aspect ratio and angle between the magnetic field and major axis of the dense cores do not correlate with the magnetic field strength in our sample (Sec.~\ref{sec:B_ab}). 
There is no preferred orientation of the major axes of the dense cores with respect to their magnetic fields regardless of the field strengths. 
This is different from the observations and simulation results of highly magnetized clumps, where tend to form filamentary structures and undergo ``aligned fragmentation'' \citep{2018A&A...615A..94F,2019ApJ...878...10T}. 
Our results are more consistent with numerical simulation incorporating turbulence, where the dense cores are more randomly oriented with respect to the magnetic fields \citep{2003ApJ...592..203G, 2004ApJ...605..800L}.

\par
No relation between magnetic fields and fragmentation has also been observed in the other star-forming regions.
The observations of clumps on a 1 pc scale in 20 high-mass star-forming regions find no correlation between the number of fragments and the magnetic field strength or mass-to-flux ratio, where the mass-to-flux ratios range from 1.8 to 7.1 \citep{2024A&A...682A..81B}. 
In contrast, in more strongly magnetized regions, the observations show the fragmentation levels are higher in dense cores ($\sim0.15$ pc) with larger mass-to-flux ratios, where the mass-to-flux ratios range from 0.2 to 1 \citep{2021ApJ...912..159P}. 
Similar tendency is also observed in core ($\sim0.3$ pc) fragmentation in hub-filament systems with stronger magnetic fields with mass-to-flux ratios of 0.3--1.6 \cite{2020A&A...644A..52A}. 
In our sample, the mass-to-flux ratios in the dense cores range from 0.3 to 6.8 (Figure \ref{fig:lambda}), and half of them are larger than 1. 
Therefore, no significant effect of magnetic fields on fragmentation could be due to the relatively weak magnetic fields being dominated over by gravity in the dense cores in Orion~A. 

\section{Summary} \label{sec:summary}
\par
To investigate the key physical parameters influencing the fragmentation process, we have assessed and compared the strengths of the turbulence, magnetic fields, and gravity of 15 fragmented and 38 unfragmented dense cores in Orion A using the JCMT POL-2, \textit{Herschel} dust temperature, and Nobeyama 45m N$_2$H$^+$ (1--0) data. 
Our results show that:
\begin{enumerate}
\item The dense cores forming binary or multiple systems tend to exhibit higher density, Mach numbers, ratios of core radii to Jeans radii, and ratios of core mass to Jeans mass, and the majority of the dense cores with binary/multiple systems have Mach numbers larger than 1.
On the other hand, there is no significant difference between the core sizes, morphologies, ratios of turbulent pressure to gravitational energy density, and mass-to-flux ratios of the dense cores with single and binary/multiple systems. 
In addition, the numbers of protostars in the dense cores tend to correlate with the density, turbulent velocity, and ratio of core size to Jeans lengths, and ratio of core mass to Jeans mass of the dense cores.
\item Our results are consistent with the turbulent core fragmentation scenario.
The supersonic turbulence likely promotes the formation of multiple density peaks in these dense cores, and these density peaks tend to locally collapse if the ratio of the core size to Jeans length (or core mass to Jeans mass) is large because of the high density of the dense core, while the turbulent support against gravity does not affect the occurrence of fragmentation in these dense cores.
\item The magnetic field does not influence the morphology and orientation of the dense cores and the occurrence of fragmentation.
This is likely because of the relatively weak magnetic field compared to gravity in these dense cores, where more than half of the dense cores have mass-to-flux ratios larger than one.
\end{enumerate}

This research has made use of data from the Herschel Gould Belt survey (HGBS) project (http://gouldbelt-herschel.cea.fr). 
The HGBS is a Herschel Key Programme jointly carried out by SPIRE Specialist Astronomy Group 3 (SAG 3), scientists of several institutes in the PACS Consortium (CEA Saclay, INAF-IFSI Rome and INAF-Arcetri, KU Leuven, MPIA Heidelberg), and scientists of the Herschel Science Center (HSC).
The James Clerk Maxwell Telescope is operated by the East Asian Observatory on behalf of The National Astronomical Observatory of Japan; Academia Sinica Institute of Astronomy and Astrophysics; the Korea Astronomy and Space Science Institute; the National Astronomical Research Institute of Thailand; Center for Astronomical Mega-Science (as well as the National Key R\&D Program of China with No. 2017YFA0402700). Additional funding support is provided by the Science and Technology Facilities Council of the United Kingdom and participating universities and organizations in the United Kingdom and Canada.
Additional funds for the construction of SCUBA-2 were provided by the Canada Foundation for Innovation.
The Nobeyama 45-m radio telescope is operated by Nobeyama Radio Observatory, a branch of National Astronomical Observatory of Japan.
This paper makes use of the following ALMA data: ADS/JAO.ALMA\#2015.1.00041. 
ALMA is a partnership of ESO (representing its member states), NSF (USA) and NINS (Japan), together with NRC (Canada), NSTC and ASIAA (Taiwan), and KASI (Republic of Korea), in cooperation with the Republic of Chile. 
The Joint ALMA Observatory is operated by ESO, AUI/NRAO and NAOJ.
H.-W.Y.\ acknowledges support from the National Science and Technology Council (NSTC) in Taiwan through grant NSTC 113-2112-M-001-035- and from the Academia Sinica Career Development Award (AS-CDA-111-M03).

\appendix
\section{Gallery of the temperature, N2H+ (1--0), and polarization data adopted in the present paper} \label{app:map}
\begin{figure*}
    \centering
    \subfigure{\includegraphics[width=0.4\textwidth]{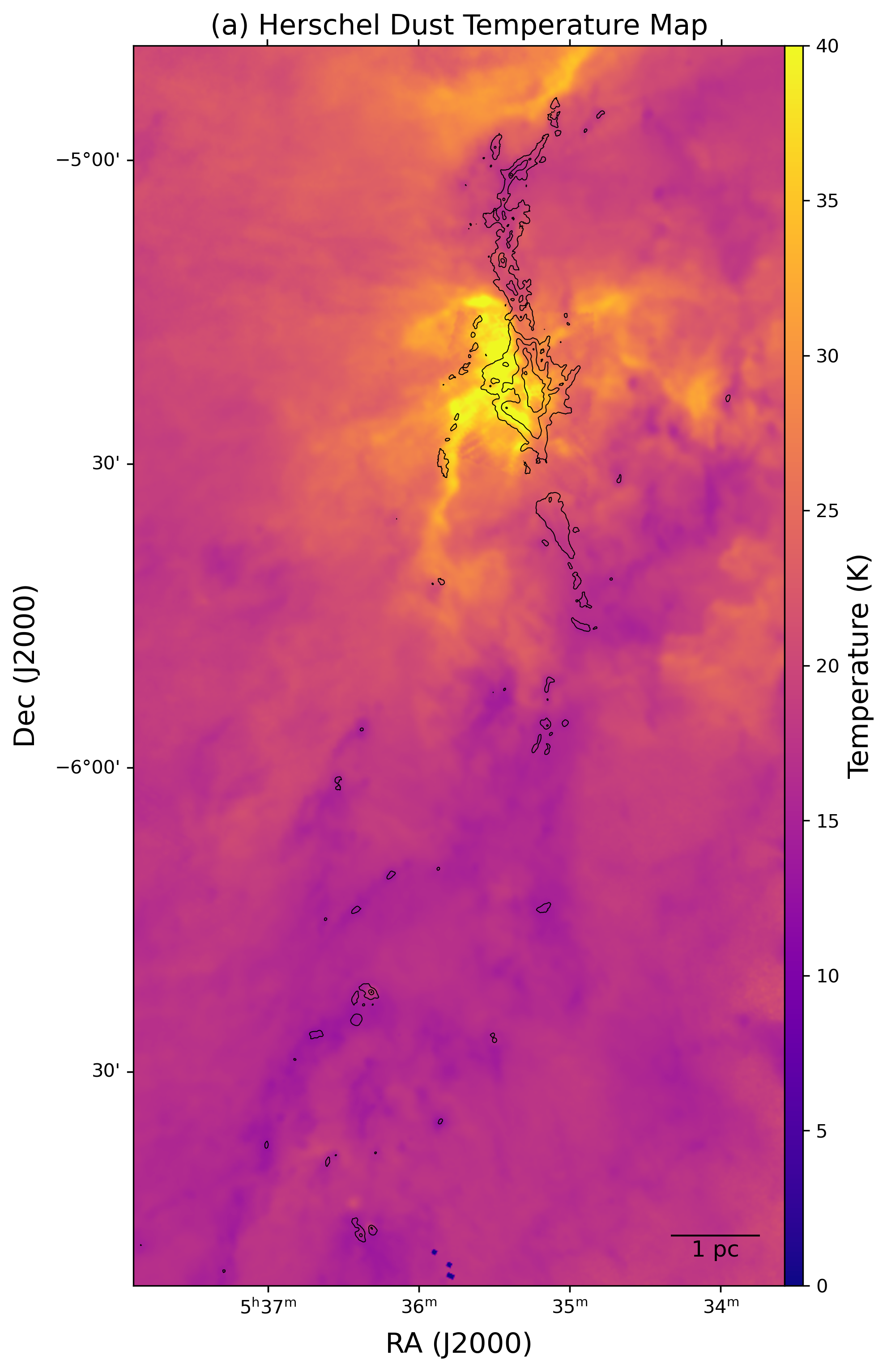}
    \label{fig:temp_map}
    }
    
    \quad
    \subfigure{\includegraphics[width=0.8\textwidth]{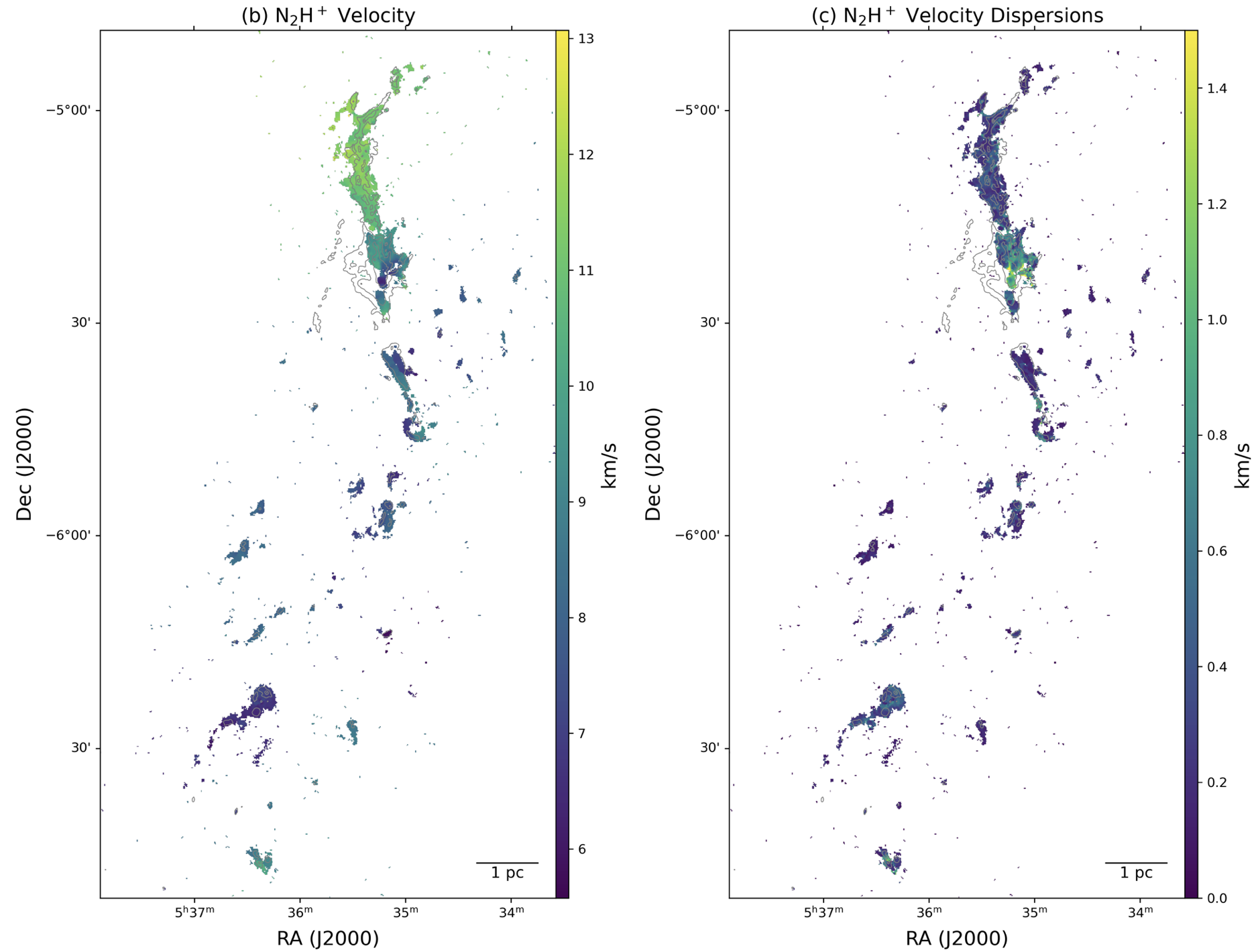}}
    \caption{(a) Herschel Dust Temperature Map and (b) N$_2$H$^+$ \textbf{(1--0)} velocity and (c) velocity dispersion map from the Nobeyama 45-m data overlaid with the JCMT 850 $\mu$m map (contours).
    The black contour levels are 0.2, 1, and 3 Jy beam$^{-1}$.}
    \label{fig:map_data}
\end{figure*}

\begin{figure*}
    \centering
    \subfigure{
        \includegraphics[width=0.45\textwidth]{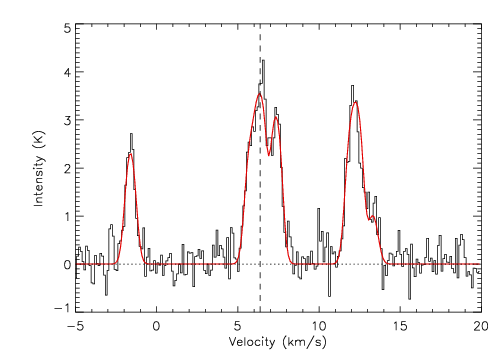}
    }
    \quad
    \subfigure{
        \includegraphics[width=0.45\textwidth]{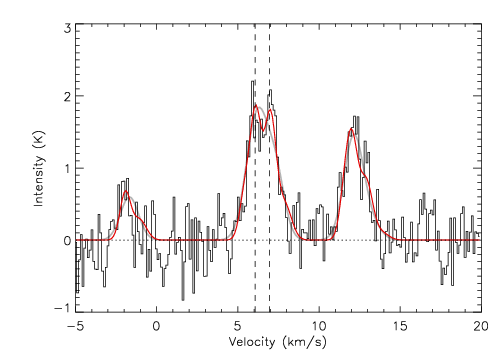}  
    }
    \caption{The examples of the N$_2$H$^+$ \textbf{(1--0)} spectra and hyperfine fitting with one and two velocity components. 
    Black histograms are the observed spectra. 
    Red lines present our best-fit spectra. 
    One and two velocity components are adopted in the left and right panels, respectively. 
    A gray line in the right present the fitting result if one velocity component is adopted for comparison.}
    \label{fig:spec}
\end{figure*}

\begin{figure*}
    \centering
    \subfigure{\includegraphics[width=1\textwidth]{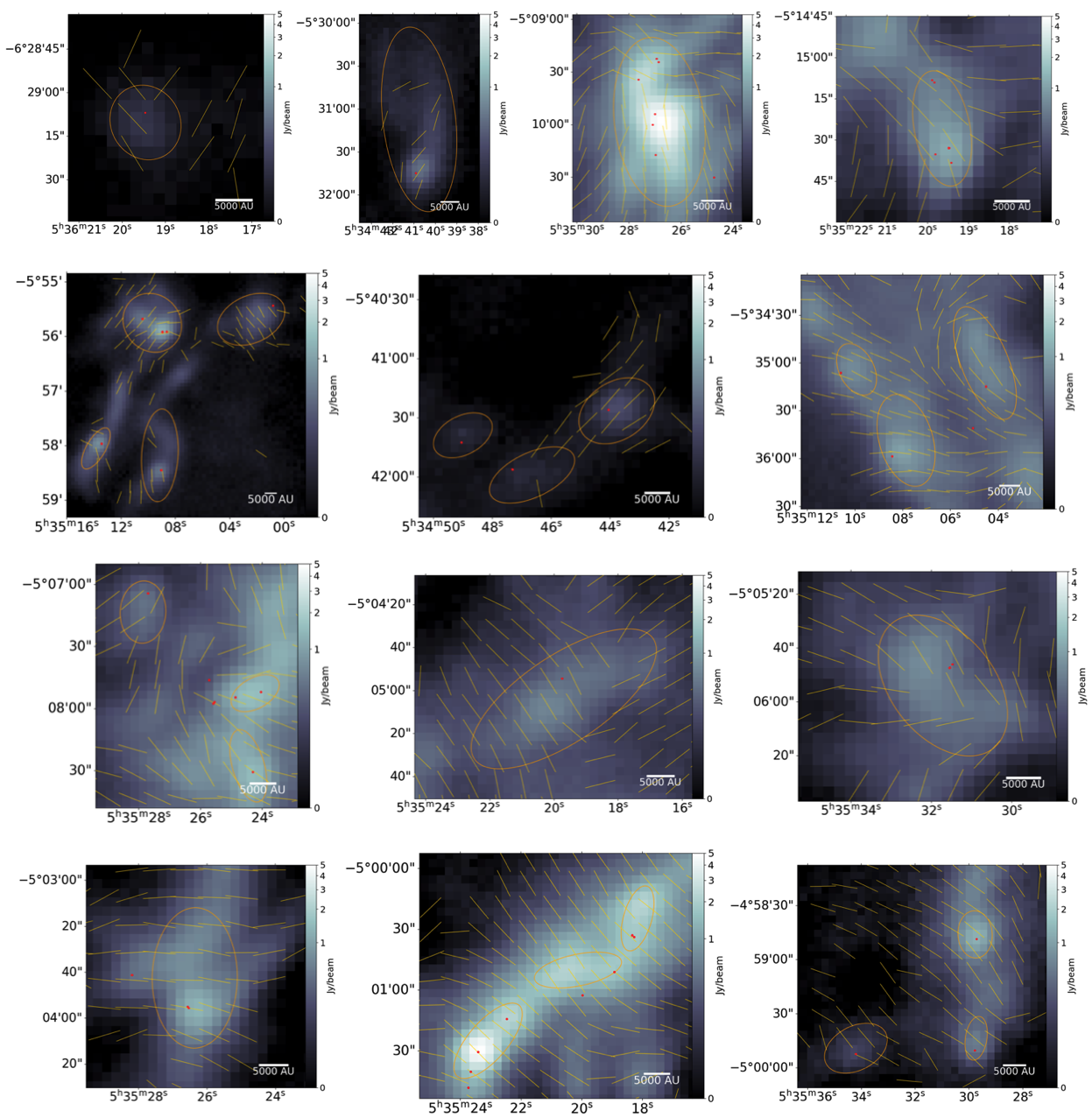}
    }
    \caption{Same as Fig.~\ref{fig:map_in} but for other dense cores in Table \ref{tab:B}.}
    \label{fig:core}
\end{figure*}

Figure \ref{fig:temp_map} shows the Herschel dust temperature map retrieved from the archive.
Figure \ref{fig:map_data}(b) and (c) present the velocity and velocity dispersion maps measured from the Nobeyama 45-m N$_2$H$^+$ \textbf{(1--0)} data.
Figure \ref{fig:spec} presents the examples of the N$_2$H$^+$ \textbf{(1--0)} spectra and our hyperfine fitting with one and two velocity components.
Details of these data are described in Sec,~\ref{sec:data}.
Figure \ref{fig:core} shows the POL-2 maps of the protostellar dense cores with polarization detections in our sample.

\newpage

\bibliography{sample631}{}
\bibliographystyle{aasjournal}

\end{document}